\documentclass[aps,superscriptaddress, showpacs,preprintnumbers, superscriptaddress, nofootinbibt,twocolumn]{revtex4-1}
\usepackage{eurosym}
\usepackage{dcolumn}
\usepackage{bm}
\usepackage{enumerate}
\usepackage{float}
\usepackage{epstopdf}
\usepackage{amsmath}
\usepackage{bm}
\usepackage{amsfonts}
\usepackage{amssymb}
\usepackage{graphicx}
\usepackage{alphalph,mathtools}
\usepackage{etoolbox}
\usepackage{color}

\def\be{\begin{equation}}
\def\ee{\end{equation}}
\def\bea{\begin{eqnarray}}
\def\eea{\end{eqnarray}}

\newcommand{\cH}{\ensuremath{\mathcal{H}}}

\begin{document}

\title{Dark energy and accelerating cosmological evolution from osculating Barthel-Kropina geometry}
\author{Rattanasak Hama}
\email{rattanasak.h@psu.ac.th}
\affiliation{Faculty of Science and Industrial Technology, Prince of Songkla University,
Surat Thani Campus, Surat Thani, 84000, Thailand,}
\author{Tiberiu Harko}
\email{tiberiu.harko@aira.astro.ro}
\affiliation{Department of Theoretical Physics, National Institute of Physics
and Nuclear Engineering (IFIN-HH), Bucharest, 077125 Romania,}
\affiliation{Department of Physics, Babes-Bolyai University, Kogalniceanu Street,
	Cluj-Napoca, 400084, Romania,}
\affiliation{Astronomical Observatory, 19 Ciresilor Street,
	Cluj-Napoca 400487, Romania,}
\author{Sorin V. Sabau}
\email{sorin@tokai.ac.jp}
\affiliation{School of Biological Sciences, Department of Biology, Tokai University, Sapporo 005-8600, Japan,}
\affiliation{Graduate School of Science and Technology, Physical and Mathematical Sciences, \\
Tokai University, Sapporo 005-8600, Japan,}

\begin{abstract}
Finsler geometry is an important extension of Riemann geometry, in which to each point of the spacetime manifold an arbitrary internal variable is associated. Interesting Finsler  geometries, with many physical applications, are the Randers and Kropina type geometries, respectively.  A subclass of Finsler geometries is represented by the osculating Finsler spaces, in which the internal variable is a function of the base manifold coordinates only. In an osculating Finsler geometry one introduces the Barthel connection, which has the remarkable property that it is the Levi-Civita connection of a Riemannian metric. In the present work we consider the gravitational and cosmological implications of a Barthel-Kropina type geometry. We assume that in this geometry the Ricci type curvatures are related to the matter energy-momentum tensor by the standard Einstein equations. The generalized Friedmann equations in the Barthel-Kropina geometry are obtained by considering that the background Riemannian metric is of Friedmann-Lemaitre-Robertson-Walker type. The matter energy balance equation is also derived. The cosmological properties of the model are investigated in detail, and it is shown that the model admits a de Sitter type solution, and that an effective dark energy component can also be generated. Several cosmological solutions are also obtained by numerically integrating the generalized Friedmann equations. A comparison of two specific classes of models with the observational data and with the standard $\Lambda$CDM model is also performed, and it turns out that the Barthel-Kropina type models give a satisfactory description of the observations.
\end{abstract}

\date{\today }
\maketitle
\tableofcontents


\section{Introduction}

Finsler geometry \cite{F1b,F2b,F3b,F4b,F5b,F6b} is a well studied generalization of the Riemann geometry, which is constructed from the general line element $ds =F\left(
x^1, x^2, ..., x^n; dx^1, dx^2, ..., dx^n\right)=F(x,y)$, where $F(x, y) > 0$, $\forall y\neq 0$, is a function on the tangent bundle $T (M)$, and
homogeneous of degree 1 in $y$. But in fact Finsler geometry is not just an extension of Riemannian geometry, but it is  Riemannian geometry without the quadratic line element restriction \cite{Chern}. If in Riemannian geometry  $F^2 = g_{IJ} (x)dx^Idx^J$ \cite{r1}, in a Finsler geometry the differential line element $ds$
at $x$ in general is a function of both $x$ and $y$, and it is defined according to $\sqrt{d\vec{x}\cdot d\vec{x}}=\left[g_{AB} (x, y) dx^Adx^B\right]^{1/2}$, respectively. In the present paper capital Latin indices take values in the range $(0,1,2,3)$, while small Latin letters take values in the range $(1,2,3)$.
Finsler spaces have a much richer mathematical structure as compared to their Riemannian counterparts. In a Finsler space one can define three kinds of curvature
tensors $\left(R^\kappa _{\nu \lambda \mu}, S^\kappa _{\nu \lambda \mu}, P^\kappa _{\nu \lambda \mu}\right)$, and five torsion tensors \cite{F5b}, which shows that indeed in Finsler geometry has a higher degree of complexity than the Riemann one.

Finsler spaces do have a large number of physical applications. Randers \cite{Rand}  unified theory of gravity and electromagnetism, initially constructed in the
framework of five dimensional general relativity, proved to be an example of a Finsler geometry, with $ds = (\alpha + \beta) du$, where $\alpha =\left[g_{IK}(x)y^Iy^K\right]^
{1/2}$, and $\beta = b_I(x)y^I$, where $y^I =dx^I/du$. Quantum mechanics in its hydrodynamical formulation can also be interpreted in terms of Finslerian geometry \cite{Tav1,Tav2,Tav3,Tav4}.

The Einstein gravitational field equations  $G_{\mu \nu} = R_{\mu \nu} - (1/2)Rg_{\mu \nu} = \kappa ^2 T_{\mu \nu}$, where $R_{\mu \nu}$ is the contraction of the Riemann curvature tensor, $R_{\mu \nu}=R^{\kappa}_{\mu \kappa \nu}$, $R$ is the Ricci scalar $R=R_{\mu}^{\mu}$, $T_{\mu \nu}$ is the matter energy-momentum tensor, and $\kappa ^2
=8\pi G/c^4$ is the gravitational coupling constant, are the mathematical representation of one of the most beautiful and successful physical theory ever proposed, general relativity. However, recently general relativity began to face a number of strong theoretical challenges. The data obtained from the observations of the Type Ia supernovae in \cite{Ri98,Ri98-1,Ri98-2,Ri98-3, Ri98-4,Hi,Ri98-5} suggests that the Universe is presently in an accelerated, de Sitter type,  expansionary phase. These observations have led to many observational and theoretical works trying to understand the present day cosmological dynamics (for a review of the cosmic acceleration problem see \cite{Ri98-6}). The Planck satellite studies of the Cosmic Microwave
Background \cite{Pl}, together with the investigations of the Baryon Acoustic Oscillations \cite{Da1,Da2,Da3} also confirmed the accelerating expansion of the Universe. But to explain these remarkable discoveries we need an essential change in our present understanding of the gravitational interaction. The simplest explanation for the accelerated expansion is obtained by reintroducing in the gravitational field equations the cosmological constant $\Lambda$, proposed by Einstein in 2017 \cite{Einb} to obtain a static model of the Cosmos. The Einstein field equations together with the cosmological constant provide very good fits to all the observational data, but with the very high price of introducing a parameter whose physical (or geometrical) nature is not (yet) known  (for detailed discussions on the cosmological constant problem see \cite{W1b,W2b,W3b}. Hence, in order to solve some of the theoretical problems of cosmology without resorting to the cosmological constant, the existence of a new essential component of the Universe, called dark energy, was postulated. There are many dark energy models that have been proposed (for extensive reviews  see \cite{PeRa03, Pa03,DE1,DE2,DE3,DE4}). Perhaps the simplest dark energy model can be constructed by using a single scalar field $\phi$, in the presence of a self-interaction potential $V(\phi)$, with the gravitational action given by
\be
S=\int{\left[\frac{M_{p}^2}{2}R-\left(\partial \phi\right)^2-V(\phi)\right]\sqrt{-g}d^4x},
\ee
where $R $ is the Ricci scalar, and $M_p$ denotes the Planck mass. The dark energy models obtained in this way are called quintessence models \cite{quint1b,quint2b,quint3b,quint4b, quint5b}.

The dark matter problem represents another fundamental problem in present day  astrophysics and cosmology (see \cite{DMR1,DMR2,DMR3} for detailed reviews of the recent results
on the search for dark matter, and for its properties). The presence of dark matter at galactic and extragalactic scales is required to obtain an explanation of two basic astrophysical/astronomical observations, the behavior of the galactic rotation curves, and the virial mass problem in clusters of galaxies, respectively. The observations of the galactic rotation curves \cite{Sal, Bin, Per, Bor} convincingly show that Newtonian gravity, as well as Einstein's general relativity, cannot describe galactic dynamics. To explain the properties of the galactic rotation curves and to solve the missing mass problem in clusters of galaxies it is necessary to postulate the existence of a dark (invisible) component of the Universe that interacts only gravitationally with baryonic matter, and which resides in a spherically symmetric halo around the galaxies. Usually, dark matter is described as a cold, pressureless cosmic fluid. Many candidates for the dark matter particle have been proposed, including WIMPs (Weakly Interacting Massive Particles), axions, neutrinos, gravitinos, neutralinos etc. (for extensive reviews of the dark matter particle candidates see \cite{Ov, Ov1,Ov2,Ov3}). Dark matter particles interact with normal baryonic matter very weakly, however, their interaction cross sections are expected to be non-zero. Thus, their direct experimental detection in laboratory experiments may be possible.

Therefore, in the present day view of the cosmological evolution, the local dynamics and the global expansion of the Universe are dominated by two major components, dark energy, and cold dark matter, respectively, with baryonic matter playing an insignificant role in the late time cosmological evolution. The simplest theoretical model, which  can fully explain the late time accelerating de Sitter expansion, is obtained by reintroducing in the Einstein gravitational field equations the cosmological constant $\Lambda$.  The $\Lambda$ extension of the Einstein gravitational field equations represents the theoretical foundation of the presently dominant standard cosmological paradigm, the $\Lambda$CDM ($\Lambda$ Cold Dark Matter) model, with cold dark matter also playing a fundamental role. Despite its theoretical simplicity, the $\Lambda$CDM model fits very well the cosmological observations \cite{C1,C2,C3, C4}.

However, besides its theoretical and fundamental aspects, the $\Lambda $CDM model is also confronted with some interesting (and not yet solved) observational problems.  The most important of these problems is the "Hubble tension", which originated from the important differences obtained between the numerical values of the Hubble constant, $H_0$, as determined by the Planck satellite from the measurements of the Cosmic Microwave Background  \cite{C4}, and the values measured directly by using cosmological and astrophysical observations in the local Universe \cite{M1,M2,M3}. To illustrate the differences, the SH0ES determinations of $H_0$ give the value
$H_0 = 74.03 \pm 1.42$ km/s/Mpc \cite{M1}, while the analysis of the CMB, originating in the early Universe, gives, via the Planck satellite data,  $H_0 = 67.4 \pm 0.5$ km/s/Mpc \cite{C3}, a value differing from the SH0ES result by $\sim  5\sigma$.

Hence, the search for alternative explanations of the cosmological dynamics, and of the nature of the two mysterious major components of the Universe is a major task for present day theoretical physics. One of the attractive possibilities for solving the gravitational puzzles is to go beyond the framework of the Riemannian geometry on which general relativity is constructed, and look for more general geometries that could describe the intricate nature of the gravitational phenomena.  In this direction one very promising candidate is Finsler geometry.

Horv\'{a}th \cite{Hor1b} and Horv\'{a}th and Mo\'{o}r \cite{Hor2b} were the first who attempted to formulate a relativistic
theory of gravitation by using Finsler geometry, with their work later extended in \cite{Hor3b} and \cite{Hor4b}, respectively. The physical and cosmological implications of the Finsler geometry have been investigated from different point of view in \cite{Fin1,Fin2,Fin3,Fin4,Fin5,Fin6,Fin7, Fin8,Fin9,Fin10,Fin11,Fin12,Fin13,Fin14,Fin15,Fin16,Fin17,Fin18,Fin19,Fin20,Fin21,Fin22,Fin23,Fin24,Fin25,Fin26,Fin27,Fin28,Fin29,Fin30,Fin31,Fin32, Fin33,Fin34,Fin35,Fin36, Fin36a, Fin37}. In particular, in \cite{Fin37} the cosmological evolution in an osculating point Barthel-Randers type geometry was considered. In this type of geometry to each point of the space-time manifold an arbitrary point vector field is associated. For the Barthel-Randers geometry the connection is given by the Levi-Civita connection of the associated osculating Riemann metric. This Finsler type geometry is assumed to describe the physical properties of the gravitational field via the standard Einstein equations, as well as the cosmological dynamics. The generalized Friedmann equations in the Barthel-Randers geometry were obtained by considering that the background Riemannian metric in the Randers line element is of Friedmann-Lemaitre-Robertson-Walker type. The matter energy balance equation was obtained, and it was investigated from the point of view of the thermodynamics of irreversible processes in the presence of particle creation. The cosmological properties of the Barthel-Randers model were investigated in detail, and it was shown that the model admits a de Sitter type solution, and that an effective cosmological constant can also be generated. Several exact cosmological solutions were also obtained, and a comparison of three specific models with the observational data and with the standard $\Lambda$CDM model was performed by fitting the observed values of the Hubble parameter. It turns out that the Barthel-Randers models give a satisfactory description of the observations.

Alternative mathematical formulations within a Finsler geometric framework are also possible. For example, in \cite{Fin29}  the cosmological implications of scalar-tensor theories that are effectively obtained from the Lorentz fiber bundle of a Finsler-like geometry were investigated. In this approach it is assumed that in the Finsler space one can define a nonlinear connection with local components $N_{\mu }^{(\alpha )}\left( x^{\nu },\phi ^{(\beta )}\right) $
, where $\left( x^{\nu },\phi ^{(\beta )}\right) $ are the local coordinates, with $x^{\nu}$, $\nu=0,...3$,  the coordinates on the local manifold, and $\phi^{(\beta)}$, $\beta =1,2$ are the coordinates on the fiber. The nonlinear connection splits uniquely the total space $T E$ into a horizontal distribution $H E$ and a vertical distribution $V E$, respectively, so that $T E=H E\oplus\oplus V E$. The metric tensor can then be constructed as
\be
G=g_{\mu \nu}(x)dx^{\mu}\otimes dx^{\nu}+v_{(\alpha)(\beta)}(x)\delta \phi ^{(\alpha)}\otimes \delta \phi ^{(\beta)},
\ee
where for $g_{\mu\nu}$ a Lorentzian signature $(-,+,+,+)$ was adopted, and with the fiber variables $\phi^{(\alpha)}$ playing the role of internal variables. The gravitational field equations are obtained from a variational principle, with the action constructed as $S=\left(1/16\pi G\right)\int{\sqrt{\left|G\right|}L_Gdx^{(N)}}$, where $dx^{(N)}=d^4x\wedge d\phi ^{(1)}\wedge d\phi ^{(2)}$. For the Lagrangian density two forms were considered, the first one being $L_G=\tilde{R}-V(\phi)/\phi$, where $\tilde{R}=R-2\Box \phi/\phi+\partial _{\mu}\phi\partial ^{\mu}\phi/2\phi ^2$ is the curvature for the case of a holonomic basis, and $V(\phi)$ is the scalar field potential. A second possible choice considered in \cite{Fin29} is $L_G=\bar{R}$, where $\bar{R}$ is the curvature of a non-holonomic basis. Matter can also be added to the gravitational action, and thus the total action becomes
\be
S=\frac{1}{16\pi G}\int{\sqrt{|{\rm det}\; g|}\phi L_Gdx^{(N)}}+\int{\sqrt{|{\rm det}\; g|}\phi L_Mdx^{(N)}},
\ee
where $L_M$ is the matter Lagrangian.  By using both choices of the action the cosmological implications of the model were investigated, and it was shown that an effective dark energyw the appearance of an effective dark energy sector, sector does appear from geometrical structure of the Finsler model. Moreover, an explicit interaction between the matter and dark energy sectors is generated. As applied to the cosmological evolution of the Universe, it was shown that one can obtain a sequence of successive matter and dark-energy dominated epochs, with the parameter of the equation of state of the dark energy  being quintessence-like, phantom-like,
or experiencing the phantom-divide crossing. An exponential de Sitter solution can also be obtained.

A generalized scalar-tensor theory obtained from vector
bundle constructions was considered in \cite{Fin36}, where its kinematic, dynamical and cosmological consequences were studied. The theory is characterized by a mathematical structure constructed over a pseudo-Riemannian space-time base manifold, together with a fibre structure with two
scalar fields. Hence, one obtains a 6-dimensional vector bundle, endowed with a non-linear connection, as the geometric arena for the description of the gravitational interaction. The geodesics, the Raychaudhuri and the field equations are obtained using both
the Palatini and the metrical approach. For the cosmological metric one adopts the expression \cite{Fin36}
\bea
G&=&-dt\otimes dt+a^2(t)\Bigg(dx\otimes dx+dy\otimes dy
+dz\otimes dz\Bigg) \nonumber\\
&&+\phi(t)\left(\delta \phi ^{(1)}\otimes \delta \phi^{(1)}+\delta \phi ^{(2)}\otimes \delta \phi^{(2)}\right),
\eea
and the corresponding Friedmann equations become \cite{Fin36}
\be
3H^2=8\pi G\left(\rho_m+\rho_{eff}\right),
\ee
\be
2\dot{H}=-8\pi G \left(\rho_m+\rho_{eff}+p_m+p_{eff}\right),
\ee
where
\be
\rho_{eff}=\frac{1}{8\pi G}\left[\frac{\dot{\phi}}{\phi}W_+-\frac{\dot{\phi}^2}{4\phi^2}-3H\left(\frac{\dot{\phi}}{\phi}-W_+\right)\right],
\ee
and
\bea
p_{eff}&=&\frac{1}{8\pi G}\Bigg[\left(W_+\right)^2-\dot{W}_+-2HW_+-\frac{\dot{\phi}^2}{4\phi ^2}\nonumber\\
&&+\frac{1}{2\phi}\left(4H\dot{\phi}-3W_+\dot{\phi}+\ddot{\phi}\right)\Bigg],
\eea
respectively, where $W_+$ is the contribution coming from the non-diagonal components of the field equations. Hence, the Finsler geometric structure generates extra terms in the modified Friedmann equations, corresponding to an effective dark energy sector, also leading to an interaction of the dark matter sector with the metric. The parameter of the dark energy equation of state can take values either in the quintessence or in the phantom regimes, or indicating the phantom-divide crossing.

It is the goal of the present paper to extend the investigations of the cosmological properties of the Finsler spaces, as initiated in \cite{Fin37}, by considering  a systematic investigation of another Finsler type geometry, which is based on the Kropina metric \cite{Krop,Krop1}, a particular class of the general $(\alpha, \beta)$ metrics with $F=\alpha ^2/\beta$. The Randers metric also belongs to the class of $(\alpha, \beta)$ metrics. We assume that the true description of the gravitational interaction can be done by using the mathematical properties of the Kropina spaces, in a mathematical framework introduced by Barthel \cite{Ba1,Ba2},  in which one can consider a Finsler space as an $n$-dimensional point space, which is locally Minkowskian (a space that is flat, homogeneous, and anisotropic) but, in general, it is not locally Euclidean. Note that a general Finsler space is both inhomogeneous and anisotropic.

The starting point of the theory of the gravitational interaction based on the point Finsler spaces is the assumption that the gravitational field can be represented by a Riemannian metric $g(x)$, satisfying the Einstein gravitational field equations. The next step in the construction of a Finslerian type gravitational theory is the non-localization (anisotropization) of the background geometry, by attaching to each
point $x = \left(x^I\right)$, $I = 0, 1, 2, 3$, {\it an internal variable} $y= \left(y^J\right)$, $J = 0, 1, 2, 3$. Under the assumption that $y$ is a vector, the nonlocal gravitational field can be described by using a Finsler type geometry $F^4$, or, alternatively,  by the geometry of a general vector bundle. {\it In the nonlocal Finslerian geometry the metric tensor depends on both the local coordinates} $x$ {\it and on the vector} $y$, so that $\hat{g} = \hat{g}(x, y)$.

However, for many realistic physical processes one can assume that the variable $y$ is a function of the position, $ y = Y (x)$.  Thus, the Finslerian metric becomes
$\hat{g} = \hat{g} \left(x, Y (x)\right))$. Hence, in this particular type of physical models the Finslerian metric tensor $\hat{g}$b is a function of $x$ only. In this way we define a manifold called the osculating Finsler manifold. The geometric properties of the osculating Finsler space are described by using the Barthel connection \cite{Ba1,Ba2}. In the case of the Kropina geometry (as well as in all $(\alpha, \beta)$ geometries) we have the remarkable result that the Barthel connection is the Levi-Civita connection of the Riemannian metric $\hat{g} (x, y(x))$. In the present work we investigate in a systematic and rigorous way the cosmological implications of the Barthel-Kropina geometry, by adopting for the background Riemann metric the Friedmann-Lemaitre-Robertson-Walker (FLRW) form. In the  Barthel-Kropina-FLRW geometry we obtain the generalized Friedmann equations of the model. A detailed investigation of the properties of the cosmological equations is performed,
and several evolutionary scenarios are constructed, describing different types of dynamics.  The model admits the de Sitter type solution, describing an accelerating Universe, but in the presence of a non-zero matter density. Different cosmological solutions corresponding to some particular choices of the coefficients of the one form $\beta$ are also considered, and their behavior is compared with the observational data, as well as with the predictions of the standard $\Lambda$CDM model.
An effective dark energy model can also be generated from the generalized Friedmann equations, and its properties are analyzed in detail.  Decelerating solutions can also be obtained, and they may provide, for example,  alternatives descriptions to the radiation dominated phase of standard cosmology. The overall comparison of the considered  models with the observational data indicate that they provide a satisfactory description of the cosmological dynamics, at least on a qualitative level.

The present paper is organized as follows. In Section~\ref{sect1} we review the necessary elements and definitions of the Finsler geometry, point and osculating Finsler spaces. The Barthel connection is also introduced in an intuitive way. The general definitions of the curvature tensors are also presented, and the specific form of the Einstein gravitational field equations is postulated. The cosmological framework involving the definitions of the Riemann metric, of the one form $\beta$, as well as of the thermodynamic parameters is constructed in Section~\ref{sect2}.  The Finslerian metric tensor coefficients, the Christoffel symbols, and the curvature tensors are obtained for the considered geometry. Based on this geometric results the generalized Friedmann equations, describing the cosmological evolution in the Barthel-Kropina-FLRW geometry are derived. Particular cosmological models are considered in Section~\ref{sect3}. It is shown that the model admits a de Sitter type solution, in the presence of a non-zero matter density. Cosmological models corresponding to two specific choices of the function $\eta$, giving the only non-zero coefficient of the one-form $\beta$, are considered in detail. The comparison of the predictions of the model with a specific dependence on the Hubble function of $\beta$ is compared with observational data  and the standard $\Lambda$CDM model. The possibility of obtaining an effective dark energy model, equivalent with a time varying cosmological constant, is analyzed in Section~\ref{sect4}. We discuss and conclude our results in Section~\ref{sect5}. The calculations of all the geometric quantities is presented in five Appendices. Thus, in Appendix~\ref{app1} the details of the computation of the Finsler metric tensor are explicitly shown. The expressions of the Christoffel symbols in the Barthel-Kropina-FLRW geometry are derived in Appendix~\ref{app2}. The components of the Ricci tensors are obtained in Appendix~\ref{app3}. The Ricci scalar is computed in Appendix~\ref{app4}. Finally, the explicit computations leading to the Einstein gravitational field equations, and to the generalized Friedmann equations, are presented in Appendix~\ref{app5}.

\section{A brief review of the Finsler, $(\alpha, \beta)$, and Barthel-Kropina geometries}\label{sect1}

In the present Section we will briefly introduce the basics of the Finsler geometry, of the $(\alpha,\beta)$-metrics, the Barthel connection, as well as the Barthel-Kropina geometry.

It is important to emphasize that Finsler geometry already appears in classical Newtonian mechanics when dissipative effects are present. If the
equations of motion of a dynamical system, defined on an $n$-dimensional differentiable manifold $M$, can be obtained from a regular Lagrangian $%
L$ via the Euler-Lagrange equations, given by
\begin{equation}
\frac{d}{dt}\frac{\partial L}{\partial y^{i}}-\frac{\partial L}{\partial
x^{i}}=F_{i},i\in \left\{1,2,...,n\right\},  \label{EL}
\end{equation}
where $F_{i}$ are the external forces, the above Euler-Lagrange
equations are equivalent to a system of second-order differential equations
of the form
\begin{equation}
\frac{d^{2}x^{i}}{dt^{2}}+2G^{i}\left( x^{j},y^{j},t\right) =0,i\in \left\{
1,2,...,n\right\} ,  \label{EM}
\end{equation}
where each function $G^{i}\left(
x^{j},y^{j},t\right) $ is $C^{\infty }$ in $\Omega $ in a neighborhood of some initial conditions $\left( \left( x\right)
_{0},\left( y\right) _{0},t_{0}\right) $. Eqs.~(\ref{EM})  can naturally be interpreted as describing geodesic motion in a Finsler space.

\subsection{Finsler geometry, and particular Finsler spaces}

A basic assumptions of the present day theoretical physics is that
space and time are unified in a single structure (the space-time) that can be described mathematically as a four dimensional
differentiable manifold $M$, endowed with a pseudo-Riemannian metric tensor $g_{I
J} $, where $I,J,K...=0,1,2,3$. On the world line of a standard clock the space-time interval between two
events $x^{I}$ and $x^{I} + dx^{I}$  is given, according to the chronological hypothesis, by $ds=\left(g_{I
J}dx^{I}dx^{J}\right)^{1/2}$  \cite%
{Tava,Tava1}. But from a mathematical point of view one can go beyond the Riemannian mathematical structures of the space-time manifold. One of the most important metrical extensions of the Riemann geometry is the geometry introduced by Finsler \cite{F1b,F2b,F3b,F4b,F5b, F6b}.

 From a general mathematical point of view Finsler spaces are metric spaces in which the interval $ds$ between two nearby points $x=(x^{I})$ and $x+dx=(x^{I} + dx^{I})$ is given by
\begin{equation}  \label{dsF}
d\hat{s}=F\left(x,dx\right),
\end{equation}
where $F$, the Finsler metric function, is positively homogeneous of degree
one in $dx$, and thus has the property
\begin{equation}
F\left(x,\lambda dx\right)=\lambda F\left(x,dx\right).
\end{equation}

To permit the use of local coordinates in mathematical computations, the Finsler
metric function $F$ is usually written in terms of the
canonical coordinates of the tangent bundle $(x,y)=(x^I,y^I)$, where $y=y^I\dfrac{\partial}{%
\partial x^I}$, is any  tangent vector $y$ at $x$. Then the Finsler metric tensor $%
\hat{g}_{I J}$ is defined as
\begin{equation}\label{Hessian mat}
\hat{g}_{I J}\left(x,y\right)=\frac{1}{2}\frac{\partial ^2F^2\left(x,y\right)}{%
\partial y^{I}\partial y^{J}},
\end{equation}
thus allowing to write Eq.~(\ref{dsF}) as $d\hat{s}^2=\hat{g}_{I J}\left(x,y\right)y^{I}y^{J}$.
Riemann spaces are particular cases of
Finsler spaces, corresponding to $\hat{g}_{IJ}\left(x,y\right)=g_{IJ
}\left(x\right)$, $y^{I}=dx^{I}$, and $ds^2=g_{IJ}(x)dx^Idx^J$, respectively.

Another important geometric quantity, the Cartan tensor $\hat{C}(x,y)$ is defined according to
\begin{equation}
\hat{C}_{IJK}=\frac{1}{2}\frac{\partial \hat{g}_{IJ}\left( x,y\right) }{\partial
{y}^{K}}.
\end{equation}

\subsubsection{Randers, Kropina and general $(\alpha , \beta )$ metrics}

The Randers space, which has many applications in physics,  is a special kind of Finsler space, \cite{Rand}, having the Finsler metric function given by
\begin{equation}
F=\left[ g_{I J }(x)dx^{I }dx^{J }\right] ^{1/2}+A_{I }(x)dx^{I }=\alpha +\beta,
\end{equation}%
where $g_{I J }(x)$ is the fundamental tensor (metric) of a {\it Riemannian space},
and $A_{I}(x)dx^{I }$ is a linear $1$-form on the tangent bundle $TM$.  Kropina \cite{Krop,Krop1} considered Finsler spaces having metrics given by
\begin{equation}
F\left( x,y\right) =\frac{g_{IJ }(x)y^{I }y^{J }}{A_{I }(x)y^{I }}.
\end{equation}

These early results were generalized by Matsumoto \cite{Mat,Mat1}, who introduced {\it the notion
of the} $(\alpha ,\beta )$ {\it metrics}. A Finsler metric function $%
F(x,y)$ is called an $(\alpha ,\beta )$ metric when $F$ is a positively
homogeneous function $F(\alpha ,\beta )$ of first degree in two variables $%
\alpha \left( x,y\right) =\left[ g_{I J }(x)dx^{I }dx^{J }\right] ^{1/2}$
and $\beta \left( x,y\right) =A_{I }(x)y^{I }$, respectively.

In the following we will assume that $\alpha $ is {\it a Riemannian metric}, that is, it has the basic properties of being
non-degenerate (regular), and positive-definite. The Randers and the Kropina metrics
belongs to the class of the $(\alpha ,\beta )$ metrics, with $F=\alpha +\beta $ for the case of the Randers metric, and $F=\dfrac{\alpha^2}{\beta}$ for the case of the Kropina metric. Moreover, we
can introduce general $(\alpha ,\beta )$ metrics having the form
\be\label{Fab}
F(\alpha
,\beta )=\alpha \phi \left( \frac{\beta }{\alpha} \right) =\alpha \phi \left(
s\right),
\ee
where $ s=\beta /\alpha $, and $\phi =\phi (s)$ is a $%
C^{\infty }$ positive function on an open interval $(-b_{o},b_{o})$.

Denoting $L=F^{2}/2$, we obtain
for the fundamental metric tensor of the $(\alpha, \beta)$ space the expression
\bea
\hat{g}_{IJ}(x,y)&=&\frac{L_{\alpha }}{\alpha }h_{IJ}+\frac{L_{\alpha \alpha }%
}{\alpha ^{2}}y_{I}y_{J}+\frac{L_{\alpha \beta }}{\alpha }\left(
y_{I}A_{J}+y_{J}A_{I}\right) \nonumber\\
&&+L_{\beta \beta }A_{I}A_{J},
\eea
where
\be
h_{IJ}=\alpha \frac{\partial ^2\alpha (x,y)}{\partial y^I \partial y^J}=g_{IJ}-\frac{y_Iy_J}{\alpha ^2},
\ee
and the indices $\alpha $, $\beta $ of $L$ indicate partial differentiation
with respect to $\alpha $ and $\beta$. Alternatively, one can compute the components of the Finslerian metric tensor by using the formula obtained in \cite{BCS2007}. Let $\alpha=\sqrt{\epsilon g_{IJ}(x)y^Iy^J}$, $\beta=A_I(x)y^I$, where $\epsilon =\pm 1$, and $F$ given by Eq.~(\ref{Fab}). Then the Hessian $\hat{g}$ can be obtained as
\be
\hat{g}_{IJ}=\epsilon \rho g_{IJ}+\rho_0b_Ib_J+\rho_1\left(b_I\alpha_J+b_J\alpha_I\right)-s\rho_1\alpha_I\alpha_J,
\ee
where $\alpha_I:=\alpha_{y^I}$ and
\be
\rho=\phi^2-s\phi\phi',
\ee
\be
 \rho_0=\phi\phi''+\phi'\phi',
 \ee
 \be
 \rho_1=-s(\phi\phi''+\phi'\phi')+\phi\phi'.
\ee

\subsection{The Barthel connection and the osculating Finsler spaces}

In the following we will briefly review the fundamental mathematical concepts on which the gravitational applications of the Kropina geometries are based, namely, the Barthel connection, and the osculation of Finsler spaces.

\subsubsection{The Barthel connection}

Let now assume that $(M,F)$ is a Finsler space, and $Y(x)\neq 0$ is a vector field defined on $M$.
We can introduce now a specific  structure $(M^n,F(x,y),Y(x))$ representing a Finsler space %
$(M^n,F(x,y))$ {\it having a tangent vector field $Y(x)$}. {\it If $Y$ is nowhere vanishing, the Finslerian metric $\hat{g}(x,y)$ gives rise to the $Y$-Riemann metric $%
\hat{g}_{Y}(x)=\hat{g}(x,Y)$.}

An important class of Finsler spaces are the point Finsler spaces. Based on the definition introduced by Barthel \cite{Ba1,Ba2}, {\it we consider in the following
a Finsler space as an} $n$-{\it dimensional point space, which is locally Minkowskian, and, in general, not locally Euclidean}.  A general Finsler space is both inhomogeneous and anisotropic, while a Minkowski space is flat, homogeneous, but anisotropic. Accordingly, {\it we shall call the Finsler} $n${\it -space a Barthel-Finsler space, or, for short, a point Finsler space}.

Given a point vector field $Y^I(x)$ and a Finsler metric tensor $\hat{g}(x,y)$, one can construct the absolute differential of $Y$ according to the definition \cite{In1}
\be\label{DY}
DY^I=dY^I+Y^K b^I_{KH}(x, Y) dx^H,
\ee
where $b^I_{KH}(x, Y)$ are {\it the Barthel connection coefficients}. The Barthel connection coefficients are constructed with the help of the generalized Christoffel symbols $\hat{\gamma}_{JIH}$, defined as
\be
\hat{\gamma}_{JIH}:=\frac{1}{2}\left(\frac{\partial \hat{g}_{JI}}{\partial x^H}+\frac{\partial \hat{g}_{IH}}{\partial x^J}-\frac{\partial \hat{g}_{HJ}}{\partial x^I}\right),
\ee
by writing the expressions in the second term of Eq.~(\ref{DY}) as
\be
Y^Kb^I_{KH}=Y^K\left(\hat{\gamma}^I_{KH}-\hat{\gamma}^R_{KS}Y^S\hat{C}^I_{RH}\right),
\ee
and defining the Barthel connection as \cite{In1}
\be
b^I_{KH}=\hat{\gamma}^I_{KH}-\hat{\gamma}^R_{KS}Y^S\hat{C}^I_{RH}.
\ee

 The Barthel connection {\it depends on the field to which it is applied}, leading to a situation that is very different from the properties of the connection in Riemannian geometry. For anisotropic metrics, all properties can depend on the direction, and {|\it in the case of the Barthel connection, the dependence is only on the direction of the field, and not on its magnitude}. The Barthel connection {\it is the simplest one that preserves the metric function by the parallel transport}. Moreover,  for Finsler vector fields, depending on both $x$ and $y$, it allows a natural passage to the standard Cartan  geometry of Finsler spaces. Therefore, we can consider the Barthel connection {\it as the connection of a point Finsler space}.
 
 From the general theory of the Finsler connections \cite{F5b,F6b} it follows that these connections, unlike the usual Levi-Civita connection of a Riemannian metric, or, more general affine connections, do not live on the base manifold $M$, but on the total space of the tangent bundle. This essential difference sometimes induces major dissimilarities in the geometrical theory of Riemannian and Finsler manifolds.

In order to fill in this gap, one of the possibilities is to evaluate the Finsler connections on a nowhere vanishing tangent vector field $(x,Y(x))$, provided the topology of the base manifold $M$ allows the existence of such vector fields. By evaluating the fundamental tensor $g_{ij}(x,y)$ of $(M,F)$ at $(x,Y(x))$, one obtains a Riemannian metric $g_Y$ on $M$, with its own Levi-Civita connection. Likewise, by evaluating the Cartan connection of $(M,F)$ at  $(x,Y(x))$ one obtains an affine connection on $M$, in general different from the Levi-Civita connection of $g_Y$, which is the Barthel connection already introduced above. W. Barthel himself arrived at this connection studying the parallel transport on $M$.

\subsubsection{The osculating Riemannian metric}

In this Subsection we will introduce the osculating Riemannian metric associated to a Finsler metric $(M,F)$. The concept of osculating Riemannian spaces of Finsler spaces was introduced by Nazim \cite{Naz}, and it was later extensively studied in \cite{Varga}. In simple terms, the osculation means that to a rather complicated structure (Finsler geometry, Finsler connection), a simpler structure (a Riemannian metric, an affine or linear connection) is associated, with the simpler structure assumed to approximates the former in some sense. Based on this link, one can obtain results on the more complicated structure.

If we fix such a local section $Y$ of $\pi_M:TM\to M$, all geometrical objects defined on the manifold $TM$ can be pulled back to $M$. Since $\hat{g}_{IJ} \circ Y$ is a function on $U$, we can define
\begin{equation}\label{Y-riem}
 \hat{g}_{IJ}(x):=\hat{g}_{IJ}(x,y)|_{y=Y(x)},\quad x\in U.
\end{equation}

The pair $(U,\hat{g}_{IJ})$ is a Riemannian manifold, while $\hat{g}_{IJ}(x)$ is called the $Y$-{\it osculating Riemannian metric} associated to $(M,F)$.

For the osculating Riemannian metric \eqref{Y-riem} the Christoffel symbols of the first kind are defined according to
\begin{equation*}
\begin{split}
  \hat{\gamma}_{IJK}(x)&:=\frac{1}{2}\left(\frac{\partial}{\partial x^J}\left[\hat{g}_{IK}(x,Y(x))\right]
+\frac{\partial}{\partial x^K}\left[\hat{g}_{IJ}(x,Y(x))\right]\right.\\
&\left.-\frac{\partial}{\partial x^I}\left[\hat{g}_{JK}(x,Y(x))\right]\right).
\end{split}
\end{equation*}

By using the derivative law of composed functions we obtain
\begin{equation}\label{Christ for g_Y}
\begin{split}
\hat{\gamma}_{IJK}(x)&=\left.\hat{\gamma}_{IJK}(x,y)\right|_{y=Y(x)}\\
& +2\left.\left(\hat{C}_{IJL}\frac{\partial Y^L}{\partial x^K}+\hat{C}_{IKL}\frac{\partial Y^L}{\partial x^J}-\hat{C}_{JKL}\frac{\partial Y^L}{\partial x^I} \right)\right|_{y=Y(x)}.
\end{split}
\end{equation}

In the case of a general $(\alpha, \beta)$ metric for the Cartan tensor we obtain
\begin{equation}
2\hat{C}_{IJK}=\frac{L_{\alpha \beta }}{\alpha }\left(
h_{IJ}p_{K}+h_{JK}p_{I}+h_{KI}p_{J}\right) +L_{\beta \beta \beta
}p_{I}p_{J}p_{K},  \label{Cijk}
\end{equation}
respectively, where we have denoted
\begin{equation}
\begin{split}
y_{I}&=g_{IJ}y^{J},p_{I}=A_{I}-%
\frac{\beta }{\alpha ^{2}}y_{I}.
\end{split}
\end{equation}

Hence, if $Y$ is a non-vanishing global section of $TM$, so that $Y(x)\neq 0$,  $\forall x\in M$, one can always define the osculating Riemannian manifold $(M,\hat{g}_{ij})$.

In the case of an $(\alpha,\beta)$ metric, let us consider the vector field $Y=A$, where $A^{I}=g^{IJ}A_{J}$. Since the vector field $A$ is globally non-vanishing on $M$, it follows that $\beta $ has no zero points. Hence, we can introduce {\it the} $A$-{\it osculating Riemannian manifold} $(M,\hat{g}_{IJ})$, where $\hat{g}_{IJ}(x):=\hat{g}_{IJ}(x,A)$.

Let $\tilde{a}$ be the length of $A$ with respect to $\alpha $. Hence
\be
\tilde{a}^{2}=A_{I}A^{I}=\alpha ^{2}\left( x,A\right) , Y_{I}\left( x,A\right)
=A_{I}.
\ee
 Therefore the $A$-osculating Riemannian metric becomes
\begin{equation}
\begin{split}
   \hat{g}_{IJ}\left( x\right) & =\left.\frac{L_{\alpha }}{\tilde{a}}\right|_{y=A(x)}g_{IJ}\\
   & +\left.\left(
\frac{L_{\alpha \alpha }}{\tilde{a}^{2}}+2\frac{L_{\alpha \beta }}{\tilde{a}}%
+L_{\beta \beta }-\frac{L_{\alpha }}{\tilde{a}^{3}}\right)\right|_{y=A(x)} A_{I}A_{J}.
\end{split}
\end{equation}

Moreover, we have
\be
\beta \left( x,A\right) =\tilde{a}^{2}, p_{I}\left(x,A\right) =0,
\ee
and consequently from  Eq.~(\ref{Cijk}) we obtain the important result that $%
\hat{C}_{IJK}\left( x,A\right) =0$.
On other hand, for  $Y=A$, we obtain
\be
\hat{\gamma}_{IJK}(x)=\left.\hat{\gamma}_{IJK}(x,y)\right|_{y=A(x)}.
\ee

Hence, for {\it a
Finsler space with $(\alpha ,\beta )$-metric the linear $A$-connection with} $A^I=(g^{IJ}A_{J})$ {\it (the Barthel
connection), is the Levi-Civita connection of the} $A${\it -Riemannian metric}.

\subsection{The curvature tensor, and its contractions}

The curvature tensor of an affine connection with local coefficients $\left(\Gamma _{BC}^A(x)\right)$ is given by
\begin{equation}
R^A_{BCD}=\dfrac{\partial \Gamma^A_{BD}}{\partial x^C}-
    \dfrac{\partial \Gamma^A_{BC}}{\partial x^D}+\Gamma^E_{BD}\Gamma^A_{EC}
    -\Gamma^E_{BC}\Gamma^A_{ED},
    \end{equation}

The Barthel connection with local coefficients $\left(b_{BC}^A(x)\right)$ {\it is an affine connection, and hence its curvature tensor must be given by the above formula},  with $\left(\Gamma _{BC}^A(x)\right)=\left(b_{BC}^A(x)\right)$. As we have already mentioned, in the case of the Kropina metric $F=\alpha^2/\beta$, the Barthel connection coincides with the Levi-Civita connection of the osculating metric $\hat{g}_{AB}(x)=g_{AB}\left(x,A(x)\right)$, where $A_I(x)$ are the components of $\beta$, and $g_{AB}$ is the fundamental tensor of $F$. Hence, since $b_{BC}^A=\hat{\gamma}_{BC}^A$, where $\hat{\gamma}_{BC}^A$ are the Levi-Civita coefficients, we obtain for the curvature tensors the expressions
\begin{equation}
\hat{R}^A_{BCD}=\dfrac{\partial \hat{\gamma}^A_{BD}}{\partial x^C}-
    \dfrac{\partial \hat{\gamma}^A_{BC}}{\partial x^D}+\hat{\gamma}^E_{BD}\hat{\gamma}^A_{EC}
    -\hat{\gamma}^E_{BC}\hat{\gamma}^A_{ED},
    \end{equation}
    and
    \begin{equation}
\hat{R}_{BD}=
    \displaystyle\sum_A\left[\dfrac{\partial \hat{\gamma}^A_{BD}}{\partial x^A}-\dfrac{\partial \hat{\gamma}^A_{BA}}{\partial x^D}
    +\sum _E\left(\hat{\gamma}^E_{BD}\hat{\gamma}^A_{EA}-\hat{\gamma}^E_{BA}\hat{\gamma}^A_{ED}\right)\right],
 \end{equation}
respectively, where $A,B,C,D,E\in \{0,1,2,3\}$. For further details on the definitions of the affine connections and of the curvature tensors see \cite{Fin37}, and references therein.

The contractions of the curvature tensor lead to the generalized Ricci
tensor, and Ricci scalar, respectively, given by
\begin{equation}
\hat{R}_{BD}=\hat{R}_{BAD}^{A},\quad\hat{R}_{D}^{B}=\hat{g}^{BC}\hat{R}_{CD},
\end{equation}%
and
\begin{equation}
\hat{R}=\hat{R}_{B}^{B},
\end{equation}%
respectively.

\section{Cosmological evolution in Barthel-Kropina geometry}\label{sect2}

In the present Section we will consider the general framework of the cosmological evolution in a Barthel-Kropina type geometry. We will begin by formulating a theoretical framework for the geometric and physical quantities of our model, in which the geometric and physical quantities are defined. As a next step we will consider the computation of the geometric quantities (metric, Christoffel symbols and curvatures), in the adopted geometry. Finally, the generalized Friedmann equations describing the cosmological evolution are obtained.

\subsection{Metric and thermodynamic quantities}

In order to construct in a systematic way the cosmology of the Barthel-Kropina
geometry, and to facilitate the physical interpretation of the results, we adopt the following assumptions.

\paragraph {\it The Riemannian metric $g_{IJ}(x)$ in $\alpha$
is given by the Friedmann-Lemaitre-Robertson-Walker (FLRW) metric},
\be
ds^2=\left(dx^0\right)^2-a^2\left(x^0\right)\left[\left(dx^1\right)^2+\left(dx^2\right)^2+\left(dx^3\right)^2\right],
\ee
where we have adopted a coordinate system with $\left(x^0=ct,x^1=x,x^2=y,x^3=z\right)$, and $a\left(x^0\right)$ is the cosmological scale factor. The FLRW metric describes a homogeneous and isotropic Universe, in which the cosmological time $t$ is flowing uniformly.

\paragraph{The Barthel-Kropina metric components are functions of $x^0$ only.}

We adopt the \textit{cosmological principle}, which requires that \textit{the global and large scale physical and
geometrical properties of the Universe depend
on the cosmological time only}. The cosmological principle
requires that the components of the vector $A$ in the one-form $\beta$ are functions of the cosmological time only, $A_I=A_I\left(x^0\right)$.

\paragraph{The space-like components of $A$ vanish.}

The cosmological principle,  together with the diagonal nature of the FLRW metric
imposes another mathematical condition on the one form $\beta$. More exactly, we require that the space-like components of $A$ vanish, so that $A_1=A_2=A_3=0
$. If this condition would not be satisfied,  we could perform a spatial rotation, thus
obtaining a preferred direction, in the $x$ coordinates, for example.
But such a cosmological model would contradict the observationally well confirmed large scale spatial isotropy of the
Universe. Therefore we assume that \textit{the vector $A$
has only one time-like independent component}, $A_0\left(x^0\right)$. Hence,
we represent the 1-form field $\beta $ as
\begin{align}\label{special A_i}
(A_{I})=(a\left(x^0\right)\eta\left(x^0\right),0,0,0)=(A^{I}),
\end{align}
where $\eta \left(x^0\right)$ is an arbitrary function of time, to be determined from the gravitational field equations.

\paragraph{A frame comoving with matter does exist.}

Similarly to the standard general relativistic case, and its Riemannian geometry, we assume that in the Barthel-Kropina geometry one can introduce a {\it comoving frame}, in which the motion of all observers take place with the Hubble flow, defined by the Riemannian metric $g_{IJ}(x)$.

\paragraph{Thermodynamic properties.}

We assume that the thermodynamic properties of the cosmological matter can be described by {\it two physical quantities only, the energy density} $\rho c^2$, {\it and the thermodynamic pressure} $p$, respectively, defined in the standard way. From assumptions {\it c} and {\it d} it follows that the only non-vanishing components of the matter energy-momentum tensor $\hat{T}_{AB}$ are $\hat{T}_0^0=\rho c^2$, $\hat{T}_{00}=\hat{g}_{00}\hat{T}^0_0$, and $\hat{T}_k^k=-p$, $\hat{T}_{ii}=-\hat{g}_{ik}\hat{T}^k_i$, respectively.

\paragraph{The gravitational field equations}

 \textit{We postulate that the Einstein gravitational field equations can be
formulated in a Barthel-Kropina geometry as}
\begin{equation}
\hat{R}_{BD}-\frac{1}{2}\hat{g}_{BD}\hat{R}=\kappa ^2 \hat{T}_{BD},  \label{Eineq}
\end{equation}%
where $\kappa^2=8\pi G/c^4 $ is a constant, and $G$ and $c$ are the Newtonian gravitational constant, and the speed of light, respectively. $\hat{T}_{BD}$ is the matter
energy-momentum tensor, {\it constructed with the help of the usual thermodynamic
quantities as defined in the standard Riemann geometry, and of the Finslerian metric tensor} $\hat{g}_{BD}$.

In the present investigation we have adopted for the Einstein equations (\ref{Eineq}) the simplest possible mathematical form, guided by the analogy with standard general relativity. Moreover, the adopted field equations have a well defined Riemannian limit, corresponding to the case $\hat{g}(x,y)\rightarrow g(x)$, when we directly recover standard general relativity, without the cosmological constant term. There are many extensions of general relativity, and other forms of the field equations, inspired by these extensions, are also possible. For example, in the $f(R)$ modified gravity theory, introduced in \cite{Bu1}, the gravitational Lagrangian is an arbitrary  function of the Ricci scalar $R$. One could assume in the framework of a Finsler variational principle that the field equations follow from a Finsler type Lagrangian of the form $f\left(\hat{R}\right)$. Alternatively, one could construct the field equations in less rigorous approach directly by substituting in the standard general relativistic equations the Riemannian quantities with their Finslerian counterparts. Modified gravity theories with curvature-matter coupling \cite{e8, Xi, Xi1} could also be extended to a Finslerian geometric framework. Riemannian geometries with torsion \cite{Capr, tors} or nonmetricity  \cite{Ghil1,Ghil2,Ghil3} can also represent a source of inspiration for obtaining their Finslerian analogues.

Hence, based on the above assumptions, the metric properties of Barthel-Kropina-FLRW model are given by

\begin{enumerate}[(i)]
\item $\epsilon=1$;
\item $\left(A_I\right)=\left(a\left(x^0\right)\eta\left(x^0\right),0,0,0\right)=\left(A^I\right)$;
\item $\left(g_{IJ}\right)=\begin{pmatrix}
1 & 0 & 0 & 0 \\
0 & -a^2(x^0) & 0 & 0 \\
0 & 0 & -a^2(x^0) & 0\\
0 & 0 & 0 & -a^2(x^0)
\end{pmatrix};$
\item $\alpha|_{y=A(x)}=a(x^0)\eta(x^0)$;
\item $\beta|_{y=A(x)}=[a(x^0)\eta(x^0)]^2$;
\end{enumerate}

\subsection{Geometric quantities in Barthel-Kropina-FLRW geometry}

As a first step in constructing a geometric theory in the Barthel-Kropina geometry we need to obtain the components of the Finslerian metric tensor (the Hessian), which are given by (for the full computational details see Appendix~\ref{app1})
\be
\hat{g}_{IJ}(x,A)=\frac{1}{\alpha^2}\left(2\epsilon g_{IJ}-\frac{A_IA_J}{\alpha^2}\right).
\ee

Next, we need to obtain the Barthel-Kropina-FLRW Finsler metric, by using the geometric properties introduced in the previous Section.  We immediately obtain
\bea
\hat{g}_{00}&=&\frac{1}{\alpha^2}\left(2g_{00}-\frac{A_0^2}{\alpha^2}\right)
=\frac{1}{a^2\eta^2}\left[2-\frac{(a\eta)^2}{(a\eta)^2}\right]\nonumber\\
&=&\frac{1}{a^2\eta^2}.
\eea
For $i,j\in\{1,2,3\}$ we find
\bea
\hat{g}_{ij}&=&\frac{1}{\alpha^2}\left(2g_{ij}-\frac{A_iA_j}{\alpha^2}\right)
=\frac{1}{\alpha^2}\left(2g_{ii}-\frac{A_i^2}{\alpha^2}\right)\delta_{ij}\nonumber\\
&=&\frac{1}{a^2\eta^2}\left[2(-a^2)\right]\delta_{ij}
=-\frac{2}{\eta^2}\delta_{ij}.
\eea

Hence, we obtain the non-vanishing components of the Barthel-Kropina-FLRW metric tensor $\hat{g}_{IJ}$ as given by
\begin{equation*}
\hat{g}_{IJ}=\begin{cases}
\hat{g}_{00}&=\dfrac{1}{a^2(x^0)\eta^2(x^0)},\vspace{0.5cm}\\
\hat{g}_{ij}&=\dfrac{-2}{\eta^2(x^0)}\delta_{ij}, i,j=1,2,3.
\end{cases}
\end{equation*}

The inverse components of the  Barthel-Kropina-FLRW metric tensor $\hat{g}_{IJ}$ can be obtained as
$$
\hat{g}^{IJ}=\begin{pmatrix}
a^2\eta^2 & 0 & 0 & 0 \\
0 & -\dfrac{\eta^2}{2} &0 & 0\\
0 & 0 &  -\dfrac{\eta^2}{2} & 0\\
0 & 0 & 0 &  -\dfrac{\eta^2}{2}
\end{pmatrix}.
$$

For the Christoffel symbols associated to the Barthel-Kropina-FLRW metric we obtain (the detailed calculations are presented in Appendix~\ref{app2}),
\begin{eqnarray*}
\hat{\gamma}^0_{00}&=&-\frac{(\eta\cH+\eta')}{\eta},
\hat{\gamma}^0_{ij}=-\frac{-2a^2\eta'}{\eta}\delta_{ij},
\hat{\gamma}^i_{0j}=-\frac{\eta'}{\eta}\delta_{ij},
\end{eqnarray*}
where $\cH=\frac{a'}{a}$.

The non-vanishing components of the Ricci tensor are given by (see Appendix~\ref{app3} for the calculational details)
\be
\hat{R}_{00}=\frac{3}{\eta^2}\left[\eta\eta''+\eta\eta'\cH-(\eta')^2\right],
\ee
and
\be
\hat{R}_{ij}=\frac{2a^2}{\eta^2}\left(3(\eta')^2+\eta\eta''+\eta\eta'\cH\right)\delta_{ij},
\ee
respectively. For the Ricci scalar $\hat{R}=\hat{g}^{IJ}\hat{R}_{IJ}$  we find (see Appendix~\ref{app4})
\begin{equation}
\hat{R}=6a^2(\eta\eta''+\cH\eta\eta'-2(\eta')^2).
\end{equation}

Finally, we obtain the Einstein tensor components
$$
\hat{G}_{IJ}=\hat{R}_{IJ}-\frac{1}{2}\hat{R}\hat{g}_{IJ},
$$
as
\begin{equation}
\hat{G}_{00}=\hat{R}_{00}-\frac{1}{2}\hat{R}\hat{g}_{00}=\frac{3(\eta')^2}{\eta^2},
\end{equation}
and
\begin{equation}
\hat{G}_{ij}=\hat{R}_{ij}-\frac{1}{2}\hat{R}\hat{g}_{ij}=\frac{2a^2}{\eta^2}\left[-3(\eta')^2+2\eta\eta''+2\cH\eta\eta'\right]\delta_{ij},
\end{equation}
respectively. For the explicit calculations of these components see Appendix~\ref{app5}.

\subsection{The generalized Friedmann equations}

Once the geometric quantities in the Barthel-Kropina-FLRW geometry are known, we can write down the full set of Einstein gravitational field equations relating geometry and matter. By taking into account our assumptions about the physical nature of the cosmological quantities, the Einstein field equations $\hat{G}_{00}=\left(8\pi G/c^4\right)\hat{g}_{00}\rho$ and $\hat{G}_{ii}=-\left(8\pi G/c^4\right)\hat{g}_{ii}p$ leads to the system of the generalized Friedmann equations, given by
\be\label{Fr1}
\frac{3(\eta')^2}{\eta^2}=\frac{8 \pi G}{c^2}\frac{1}{a^2\eta ^2}\rho,
\ee
and
\be\label{Fr2}
a^2\left[-3(\eta')^2+2\eta\eta''+2\cH\eta\eta'\right]=\frac{8\pi G}{c^4}p,
\ee
respectively. By replacing the term $-3\left(\eta '\right)^2$ with the help of Eq.~(\ref{Fr1}), Eq.~(\ref{Fr2}) takes the alternative form
\be\label{Fr3}
2a\eta \frac{d}{dx^0}\left(\eta 'a\right)=\frac{8\pi G}{c^4}\left(\rho c^2+p\right).
\ee

\subsection{General relativistic limit, and the energy balance equation}

A first interesting property of the cosmological equations of the Barthel-Kropina-FLRW model is that {\it in the limit} $\eta \rightarrow \pm 1/a$, $\beta =(1,0,0,0)$, {\it the generalized Friedmann  Eqs.~(\ref{Fr1}) and (\ref{Fr2}) do reduce to the standard Friedmann equations of general relativity},
\be
\frac{3(a')^2}{a^2}=\frac{8 \pi G}{c^2}\rho,
\ee
\be
2\frac{\ddot{a}}{a}+\frac{(a')^2}{a^2}=-\frac{8\pi G}{c^4}p,
\ee
a result which is easy to check. From the Friedmann equations of standard general relativistic cosmology it follows that the matter energy density $\rho$ satisfies the conservation equation
\be
\dot{\rho}+3H\left(\rho +\frac{p}{c^2}\right)=0,
\ee
where the dot denotes the derivative with respect to the time, and $H=c\cH$. For $p=0$ the conservation equation can be immediately integrated to give
\be
\rho=\frac{\rho_0}{a^3}=\rho_0(1+z)^3,
\ee
where $\rho_0=\rho_{0b}+\rho_{0DM}$ is the present day matter density of the Universe, with $\rho_{0b}$ and $\rho_{0DM}$ denoting the present day density of the baryonic and dark matter, respectively,  and we have introduced the redshift variable, defined as $1/z=1+a$.  We introduce now the generalized matter density parameter $\Omega _m$, defined as
\be
\Omega _m=\frac{\rho}{\rho_{0}}=\frac{1}{a^3}=(1+z)^3.
\ee

By considering a three component Universe, filed with ordinary matter, dark matter and dark energy, respectively, in the standard $\Lambda$CDM model the Hubble function is given by
\be
H=H_0\sqrt{\frac{\Omega_m^{(cr)}}{a^3}+\Omega _{\Lambda}}=H_0\sqrt{\Omega _m^{(cr)}(1+z)^3+\Omega _{\Lambda}},
\ee
where $\Omega _m^{(cr)}=\Omega _b^{(cr)}+\Omega _{DM}^{(cr}$, and $\Omega_b^{(cr)}=\rho_b/\rho_{cr}$, $\Omega _{DM}=\rho_{DM}/\rho_{cr}$ and $\Omega _{\Lambda}=\Lambda/\rho_{cr}$ denote the density parameters of the baryonic matter, dark matter, and dark energy, respectively. Here $\rho_{cr}$ denotes the critical density of the Universe, defined as $\rho_{cr}=3H_0^2/8\pi G$. An important observational parameter, the deceleration parameter is defined as
\be
q=\frac{d}{dt}\frac{1}{H}-1.
\ee

In the $\Lambda$CDM model the deceleration parameter is obtained as
\be
q(z)=\frac{3(1+z)^3\Omega _m}{2\left[\Omega _{\Lambda}+(1+z)^3\Omega _m\right]}-1.
\ee

To perform a comparison between the predictions of the Barthel-Kropina-FLRW cosmological model with observations and with the $\Lambda$CDM model,
we adopt for the matter and dark energy density parameters the values  $\Omega_{DM} = 0.2589$, $\Omega_b = 0.0486$, and $\Omega _{\Lambda} = 0.6911$, respectively \cite{Pl}. Then for the total matter density parameter $\Omega _m = \Omega _{DM} + \Omega_b$ we obtain $\Omega _m = 0.3089$. The present day value of the
deceleration parameter has the value $q(0) = -0.5381$, indicating that presently the Universe is in an accelerating phase.

\subsubsection{The Riemannian limit of the Kropina metric}

We consider now the limiting case of the Finsler metric. Since in the case of the considered Kropina metric $F=\alpha ^2/\beta$, with $\alpha ^2=g_{IJ}(x)dy^Idy^j$, and $\beta =A_Idy^I$, the standard general relativistic Riemann geometry is recovered in the limit $\beta \rightarrow \alpha$, which gives $\lim_{\beta \rightarrow \alpha}F=\alpha=\sqrt{g_{IJ}(x)y^Iy^J}$. The condition $\alpha =\beta$ gives immediately $g_{IJ}y^Iy^J=A_IA_Jy^Iy^J$, or $\lim_{\alpha \rightarrow \beta}\hat{g}_{IJ}=A_IA_J$. This means that the limiting Riemann metric $g_{IJ}$ is actually an inner product, and hence it becomes a degenerate Riemann metric in the sense that $\det \left|g_{IJ}\right|=0$, that is, ${\rm rank}g_{IJ}=n-1$.

Equivalently, one can show by taking the limit $\beta \rightarrow \alpha$ in Eq.~(\ref{Kropina Hessian}), giving the metric tensor of the Barthel-Kropina geometry,  that $\lim_{\beta \rightarrow \alpha}\hat{g}_{IJ}=g_{IJ}=A_IA_J$. The fact that in the Riemannian limit the Kropina metric leads to a degenerate Riemann metric should not come as a surprise. Indeed, the original Kropina metric is degenerate at the origin of $T_xM$, and by taking the limit $\beta \rightarrow \alpha$ we obtain a degenerate Riemann metric.

This also means that in this limit the Kropina indicatrix degenerates into the Riemannian unit circle of the degenerate metric $g_{IJ}$. Since $\lim_{\beta \rightarrow \alpha}\hat{g}(x,y)=g_{IJ}(x)$, under the same limit the $Y$-osculating metric $\hat{g}\left(x,Y(x)\right)=\hat{g}(x)$ tends to $g_{IJ}(x)$, which is degenerate, as already explained.   Moreover, let us observe that the condition $g_{IJ}\rightarrow A_IA_J$ is incompatible with the choice $g_{IJ}={\rm diag}\left(1,-a^2,-a^2,-a^2\right)$ and $b_I=\left(b_0,0,0,0\right)$, respectively,  except for the case $a^2=0$, which is obviously meaningless in the present context.

\subsubsection{The energy balance equation}

We multiply now Eq.~(\ref{Fr1}) with $a^3$, and take the derivative of the resulting relation with respect to $x^0$. Thus we obtain
\be\label{en1}
\frac{8\pi G}{c^2}\frac{d}{dx^0}\left(\rho a^3\right)=15a^4a'\left(\eta '\right)^2+6a^5\eta ' \eta ''.
\ee

After multiplying Eq.~(\ref{Fr2}) with $3a^2a'$, we obtain
\be\label{en2}
\frac{8\pi G}{c^4}p\frac{d}{dx^0}a^3=-9a^4a'\left(\eta '\right)^2+6a^4a'\eta \eta ''+6a^4a'\cH \eta \eta '.
\ee

By adding Eqs.~(\ref{en1}) and (\ref{en2}) we obtain the energy balance equation in the Barthel-Kropina-FLRW cosmological model as
\bea\label{bal}
&&\frac{8 \pi G}{c^4}\left[\frac{d}{dx^0}\left(\rho c^2a^3\right)+p\frac{d}{dx^0}a^3\right]=6a^5\Big[\cH \left(\eta '\right)^2\nonumber\\
&&+\left(\eta '+\cH \eta\right)\eta ''+\cH ^2\eta \eta '\Big].
\eea

The energy balance equation can be reformulated, with the use of the generalized Friedmann equations, as
\bea
&&\frac{8 \pi G}{c^4}\left[\frac{d}{dx^0}\left(\rho c^2a^3\right)+p\frac{d}{dx^0}a^3\right]\nonumber\\
&&=6a^5\left[\frac{8\pi G}{2c^4}\left(\frac{5}{3}\rho c^2+p\right)\frac{\cH}{a^2}+\eta ' \eta ''\right].
\eea

\section{Particular cosmological models}\label{sect3}

In the present Section we will investigate the cosmological implications of the generalized Friedmann equations (\ref{Fr1}) and (\ref{Fr2}) obtained within the framework of the Barthel-Kropina-FLRW geometry.

\subsection{The de Sitter solution}

The de Sitter solution corresponds to an exponential expansion in the Riemann geometric framework, with $a\left(x^0\right)=e^{\cH _0x^0}$, where $\cH _0=\cH={\rm constant}$. The system of the generalized Friedmann equations does not admit an explicit vacuum solution, since $\rho =p =0$ implies $\eta '=0$, $\eta ={\rm constant}$, and the second Friedmann equation (\ref{Fr2}) is automatically satisfied. Hence, a de Sitter type exponential expansion can take place only in the presence of matter. Thus we assume a non-zero matter energy density, but a vanishing thermodynamic pressure for the cosmological matter, with $p=0$. In the case $\cH=\cH_0={\rm constant}$, Eq.~(\ref{Fr2}) becomes
\be\label{dSe}
-3(\eta')^2+2\eta\eta''+2\cH _0\eta\eta'=0.
\ee
By introducing a new variable $u$, defined as $\eta '=u$, we immediately obtain $\eta ''=du/dx^0=(du/d\eta)\left(d\eta /dx^0\right)=u(du/d\eta)$. Hence,  Eq,~(\ref{dSe}) becomes
\be
\frac{du}{d\eta}=\frac{3}{2}\frac{u}{\eta}-\cH _0,
\ee
with the general solution given by
\be
u(\eta)=C_1\eta ^{3/2}+2\cH _0\eta,
\ee
where $C_1$ is an arbitrary constant of integration. Therefore, from
\be
\frac{d\eta}{dx^0}=C_1\eta ^{3/2}+2\cH _0\eta,
\ee
we obtain the general solution of Eq.~(\ref{dSe}) as
\be
\eta \left(x^0\right)=\frac{4\cH _0^2 e^{2\cH _0\left(x^0+C_2\right)}}{\left[1\pm C_1e^{\cH _0\left(x^0+C_2\right)}\right]^2},
\ee
where $C_2$ is an integration constant that can be taken as zero without any loss of generality by a rescaling of the time coordinate $x^0$. Moreover, in order to obtain a physically consistent solution we will adopt the minus sign in the above equation. Hence, a de Sitter type exponential expansion of the Riemann metric corresponds to
\be
\eta \left(x^0\right)=\frac{4\cH _0^2 e^{2\cH _0x^0}}{\left[1- C_1e^{\cH _0x^0}\right]^2}.
\ee

During the de Sitter phase the matter energy density varies as
\be\label{rhodS}
\frac{8\pi G}{c^2}\rho \left(x^0\right)=\frac{192\cH _0^6e^{6\cH _0x^0}}{\left(1-C_1e^{\cH _0x^0}\right)^6}.
\ee

At the initial moment $x^0=0$, the matter energy density is
\be
\frac{8\pi G}{c^2}\left.\rho \left(x^0\right)\right|_{x^0=0}=\frac{192\cH_0 ^6}{\left(1-C_1\right)^6},
\ee
while for $x^0\rightarrow \infty$, we obtain
\be
\frac{8\pi G}{c^2}\mathrm{\lim _{x^0\rightarrow \infty}\rho \left(x^0\right)}=\frac{192\cH _0^6}{C_1^6}.
\ee

In order to obtain a monotonically decreasing matter energy density during the de Sitter phase, satisfying the condition $\left.\rho \left(x^0\right)\right|_{x^0=0}>\mathrm{\lim _{x^0\rightarrow \infty}\rho \left(x^0\right)}$, the integration constant $C_1$ must satisfy the condition $C_1>1/2$.

\subsection{Conservative cosmological evolution}

We will consider now conservative Barthel-Kropina-FLRW models, that is, models in which the matter conservation equation
\be\label{matc}
\frac{d}{dx^0}\left(\rho c^2a^3\right)+p\frac{d}{dx^0}a^3=0,
\ee
is identically satisfied. In this case the system of generalized Friedmann equations reduces to the Friedmann equations (\ref{Fr1}) and (\ref{Fr2}), together with the constraint
\be
\cH \left(\eta '\right)^2+\left(\eta '+\cH \eta\right)\eta ''+\cH ^2\eta \eta '=0.
\ee

\subsubsection{Linear barotropic fluid cosmological models}

As a first example of a conservative Barthel-Kropina-FLRW cosmological model we consider the case of a barotropic fluid dominated Universe, with $p=(\gamma -1)\rho c^2$, where $\gamma ={\rm constant}$, and $1\leq \gamma \leq 2$. In this case Eq.~(\ref{matc}) immediately gives
\be\label{56}
\rho \left(x^0\right)=\frac{\rho_0}{a^{3\gamma }\left(x^0\right)},
\ee
where $\rho _0$ is an integration constant. Then Eq.~(\ref{Fr1}) gives
\bea\label{59}
\eta '&=&-\sqrt{\frac{8\pi G \rho _0}{3c^2}}\frac{1}{a^{3\gamma/2+1}}, \\
\eta''&=&\left(\frac{3\gamma}{2}+1\right)\sqrt{\frac{8\pi G \rho _0}{3c^2}}\frac{a'}{a^{3\gamma/2+2}},
\eea
where we have assumed that $\eta $ is a monotonically decreasing function of time.

By taking into account the linear barotropic equation of state, with the use of Eq.~(\ref{Fr1}), the pressure can be obtained as
\be
p=(\gamma-1)\rho c^2=3(\gamma -1) \frac{c^4}{8\pi G}a^2 \left(\eta '\right)^2.
\ee

After substituting the above expression of the pressure in Eq.~(\ref{Fr2}) we obtain the equation
\be
2\eta \eta''+2\cH \eta \eta'=3\gamma \left(\eta'\right)^2,
\ee
which can be integrated immediately to obtain the first integral
\be
\left|\eta '\right|a=C^{-3\gamma /2}\eta ^{3\gamma /2},
\ee
where $C$ is an arbitrary integration constant. With the use of the expression (\ref{59}) for $\eta '$ we obtain
\be
\eta (a)=C\left(\frac{8\pi G\rho_0}{3c^2}\right)^{1/3\gamma}a^{-1}.
\ee
By taking the derivative of the above expression of $\eta$ with respect to $x^0$, and equating it with $\eta '$ obtained from the first Friedmann equation, as given by Eq.~(\ref{59}), we obtain for $a'$ the equation
\be
\frac{a'}{a^2}=\frac{1}{C}\left(\frac{8\pi G\rho_0}{3c^2}\right)^{\left(3\gamma -2\right)/6\gamma}\frac{1}{a^{3\gamma/2+1}},
\ee
with the general solution given by $a\left(x^0\right)\propto \left(x^0\right)^{2/3\gamma}$. Hence, conservative Barthel-Kropina-FLRW cosmological models exactly mimic their  general relativistic counterparts, describing the decelerating evolution of a matter dominated Universe satisfying a linear barotropic equation of state. For a dust Universe we obtain $a\left(x^0\right)=\left(x^0\right)^{2/3}$, while for a radiation dominated Universe the scale factor varies as  $a\left(x^0\right)=\left(x^0\right)^{1/2}$.

\subsection{Nonconservative Barthel-Kropina-FLRW cosmological models}

In the following we consider cosmological models in which the matter density and pressure do not satisfy the conservation Eq.~(\ref{matc}), and therefore, for a linear barotropic equation of state the expression of the energy density is different from the one given by Eq.~(\ref{56}). However, we will consider that matter still satisfies a linear barotropic equation of state. Hence, by eliminating the pressure from the second Friedmann equation (\ref{Fr2}) with the help of the equation of state and of the first Friedmann equation (\ref{Fr1}), we obtain
\be
\eta \eta ''+\cH \eta \eta '=\frac{3\gamma }{2}\left(\eta '\right)^2,
\ee
which can be immediately integrated to give the first integral
\be\label{67}
\left|\eta '\right|a=C\eta ^{3\gamma /2},
\ee
where $C$ is an arbitrary constant of integration, and we have used the standard mathematical relation $\int{\left(f'(x)/f(x)\right)dx}=\ln \left|f(x)\right|+C$. For the matter density we find the expression
\be\label{68}
\frac{8\pi G}{c^2}\rho=a^2\left(\eta '\right)^2=C^2\eta ^{3\gamma}.
\ee
In order to obtain a monotonically decreasing matter density, $\eta $ must be also a decreasing function of time. In terms of $\eta$ the Hubble function is obtained as
\be
\cH=\frac{3\gamma}{2}\frac{\eta '}{\eta}-\frac{\eta ''}{\eta '}.
\ee

Eqs.~(\ref{67}) and (\ref{68}) give the full solution of the Friedmann equations for the cosmological evolution in the Barthel-Kropina-FLRW geometry. Once the functional form of $\eta$ is specified, the evolution of the scale factor and of the other cosmological parameters can be immediately obtained.

\subsubsection{$\eta =\eta (a)$}

As a first example of a nonconservative Barthel-Kropina-FLRW cosmological model we consider the case in which $\eta$ is a function of the scale factor only, $\eta =\eta (a)$. Then Eq.~(\ref{67}) immediately gives the evolution equation
\be
a\frac{d\eta(a)}{da}a'=C\eta ^{3\gamma /2}(a),
\ee
which can be integrated to give
\begin{equation}
C\left( x^{0}-x_{0}^{0}\right) =\int a\frac{d\eta (a)}{da}\eta ^{-3\gamma
/2}(a)da,
\end{equation}%
where $x_{0}^{0}$ is an arbitrary integration constant.

In the following we will consider the case of a dust Universe with $\gamma =1$. The simplest possible cosmological model can be obtained for $\eta (a)=a_1a^{-n}$, where $a_1$ and $n$ are constants. Then we immediately obtain
\begin{equation}
a\left( x^{0}\right) =\left( \frac{5\sqrt{a_{1}}C}{2}\right) ^{2/\left(
2-5n\right) }\left( x^{0}-x_{0}^{0}\right) ^{2/\left( 2-5n\right) },
\end{equation}%
where we assume $5n<2$, $n<0.4$. For the Hubble function we find
\begin{equation}
H\left( x^{0}\right) =\frac{2}{2-5n}\frac{1}{x^{0}},
\end{equation}%
while for the deceleration parameter we obtain the expression $q=-5n/2$. The
Universe is in an accelerated expansionary state, with the scale factor
having a power law dependence on the time coordinate.

\subsubsection{$\eta =\eta \left(\cH \right)$}

As another particular class of Barthel-Kropina-FLRW cosmological models we consider the case in which $\eta$ is a function of $\cH$, the Hubble function, so that $\eta =\eta (\cH )$. Hence, under this assumption, Eq.~(\ref{67}) takes the form
\be\label{70}
\frac{d\eta \left(\cH\right)}{d\cH}\frac{d\cH}{dx^0}a=C^{-3\gamma /2}\eta ^{3\gamma/2}\left(\cH\right).
\ee

At this moment we introduce as the independent variable the redshift $z$, defined as $1+z=1/a$. Hence we obtain
\be
\frac{d}{dx^0}=\frac{dz}{dx^0}\frac{d}{dz}=-(1+z)\cH (z)\frac{d}{dz}.
\ee

In the redshift variable Eq.~(\ref{70}) becomes
\be\label{71}
-(1+z)^2\frac{d\eta \left(\cH (z)\right)}{d\cH (z)}\cH (z)\frac{d\cH (z)}{dz}=C^{-3\gamma /2}\eta ^{3\gamma/2}\left(\cH (z)\right).
\ee

In the following we adopt for $\eta (\cH )$ the simple form
\be
\eta (\cH )=a_1\cH ^n+b_1,
\ee
where $a_1$, $b_1$ and $n$ are constants. We also rescale the Hubble function according to $\cH = (1/c)H=(1/c)H_0h$, where $H_0$ is the present day value of the Hubble function. Moreover, we rescale the coefficient $a_1$ as $a_1\rightarrow a_1/H_0^n$. We consider the late-time evolution of the Universe, in its dust phase, with zero pressure, and therefore we take $\gamma =1$.  Hence, Eq.~(\ref{71}) takes the form
\be
-(1+z)^2h^n(z)\frac{dh(z)}{dz}=c_1\left(a_1h^n+b_1\right)^{3/2},
\ee
where we have denoted $c_1=C^{-3/2}/na_1$. The matter energy density can be obtained as
\be
\frac{8\pi G}{c^2}\rho =C^2\left(a_1\cH ^n+b_1\right)^3.
\ee

For the matter density parameter we obtain
\be
\Omega _m=\frac{\rho}{\rho_0}=C_1\left(a_1h^n+b_1\right)^3,
\ee
where we have denoted $C_1=C^2c^2\rho_c/3H_0^2\rho_0$.  

\begin{figure}[tbp!]
\centering
\includegraphics[width=8cm]{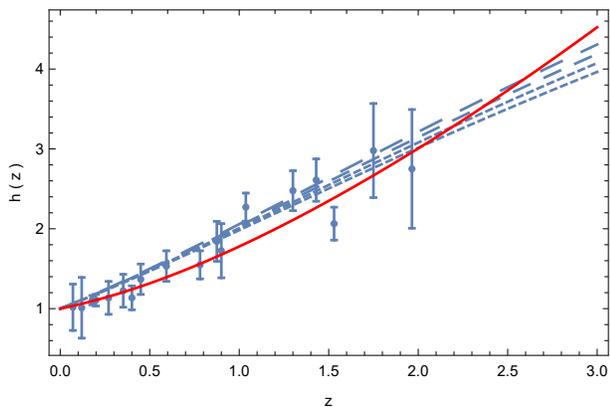}
\caption{Variation of the dimensionless Hubble function $h$ as a function of the redshift $z$ for $\eta (z)=a_1h^n(z)+b_1$ for $n=2/3$, $b_1=-0.21$,  $c_1=3.9$, and different values of $a_1$: $a_1=0.348$, (short dashed curve), $a_1=0.350$ (dashed curve), $a_1=0.352$, (long dashed curve), and $a_1=0.354$, (ultra-long dashed curve), respectively. The red solid line represents the prediction of the $\Lambda$CDM model. The observational data are represented together with their error bars.}
\label{fig1}
\end{figure}

\begin{figure}[tbp!]
\centering
\includegraphics[width=8cm]{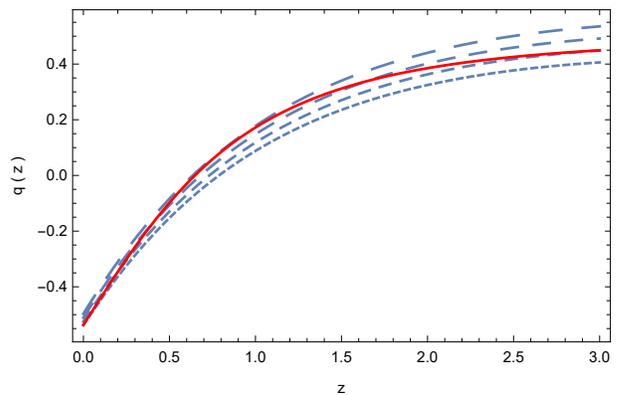}
\caption{Variation of the deceleration parameter $q$ as a function of the redshift $z$ for $\eta (z)=a_1h^n(z)+b_1$  for $n=2/3$, $b_1=-0.21$,  $c_1=3.9$, and different values of $a_1$: $a_1=0.348$, (short dashed curve), $a_1=0.350$ (dashed curve), $a_1=0.352$, (long dashed curve), and $a_1=0.354$, (ultra-long dashed curve), respectively. The red solid line represents the prediction of the $\Lambda$CDM model. }
\label{fig2}
\end{figure}

\begin{figure}[tbp!]
\centering
\includegraphics[width=8cm]{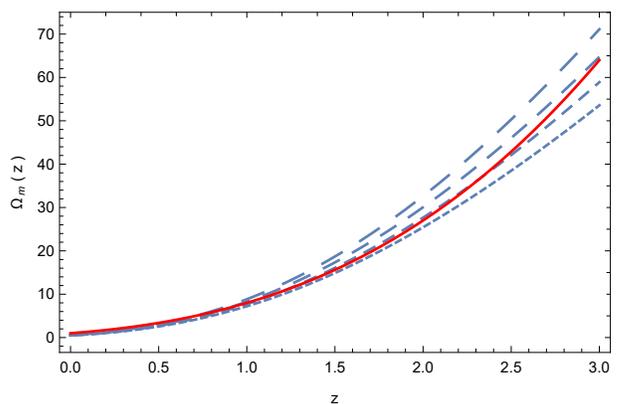}
\caption{Variation of the total matter density parameter $\Omega _m$ as a function of the redshift $z$ for $\eta (z)=a_1h^n(z)+b_1$ for $n=2/3$, $b_1=-0.21$,  $c_1=3.9$, and different values of $a_1$: $a_1=0.348$, (short dashed curve), $a_1=0.350$ (dashed curve), $a_1=0.352$, (long dashed curve), and $a_1=0.354$, (ultra-long dashed curve), respectively.  The red solid line represents the prediction of the $\Lambda$CDM model. }
\label{fig3}
\end{figure}

The variation of the normalized Hubble function, of the deceleration parameter and of the matter density parameter are presented in Figs.~\ref{fig1}-(\ref{fig3} for four values of $n$ ($n=1,2,3,4$), and for different values of $a_1$, $b_1$ and $c_1$. In the Figures we have also included the observational data for the Hubble function, together with their error bars \cite{H1,H2}, as well as the predictions of the $\Lambda$CDM model.

As one can see from Fig.~\ref{fig1}, for the adopted functional form of $\eta$, the Barthel-Kropina-FLRW model gives an acceptable description of the observational data for $h(z)$, up to a redshift of around $z\approx 3$, and it can also reproduce, at least qualitatively, the predictions of the $\Lambda$CDM model. However, with increasing $z$, at redshifts higher than 3 important deviations from the predictions of the $\Lambda$CDM model do appear, with the dimensionless Hubble functions of the $\Lambda$CDM model increasing much faster than in Barthel-Kropina-FLRW cosmology.

For the adopted set of the model parameters, the comparison of the deceleration parameter curves, presented in Fig.~\ref{fig2}, as predicted by the Barthel-Kropina-FLRW model and by the $\Lambda$CDM model, indicate the existence of significant differences between the two models, especially at higher redshifts. Both models predict a late-time accelerating behavior, with $q<0$ at the present time,  but the numerical values for $q(0)$ differ considerably in the two models at redshifts $z>0.5$. However, the present day values of $q$ in the Barthel-Kropina-FLRW cosmology are consistent with the predictions of the $\Lambda$CDM model. At larger redshifts the behavior of the deceleration parameter $q$  in the present Finsler type model strongly depends on the numerical values of the model parameters, and at redshifts higher than 3 significant deviations from the predictions of the $\Lambda$CDM model do occur.

At a qualitative level the Barthel-Kropina-FLRW cosmology can also reproduce the behavior of the total matter density parameter $\Omega_m$ in the $\Lambda$CDM model. At low redshifts $0\leq z <1$, the predictions of the two models basically agree. However, the quantitative differences, already observed in the case of the deceleration parameter, do also exist for this physical parameter.  In the range $1\leq z\leq 3$, the behavior of the matter density parameter strongly depends on the model parameters, $\Lambda$CDM, and at higher redshifts, both lower and higher matter density parameters can be obtained, thus leading to matter  densities that are different from  the predictions of the $\Lambda$CDM model.

\section{Dark energy in the Barthel-Kropina-FLRW cosmology}\label{sect4}

We will consider now that the function $\eta$ can be represented in a general form as
\be
\eta =\frac{1}{a\left(x^0\right)}\left[1+f\left(x^0\right)\right],
\ee
where $f$ is an arbitrary function to be determined from the field equations. In the limit $f\rightarrow 0$, $\eta \rightarrow 1/a$, and, as we have already seen, we recover the standard Friedmann equations of general relativity. With this representation of $\eta$ the generalized Friedmann equations take the form
\bea
3\frac{\left(a'\right)^2}{a^2}&=&\frac{8\pi G}{c^2}\rho+6(1+f)\cH f'-3\left(f'\right)^2-3\cH ^2(2+f)f\nonumber\\
&=&\frac{8\pi G}{c^2}\rho +\rho _{DE},
\eea
and
\bea
\frac{2a''}{a}+\frac{\left(a'\right)^2}{a^2}&=&-\frac{8\pi G}{c^4}\frac{p}{(1+f)^2}+4\cH \frac{f'}{1+f}-3\frac{\left(f'\right)^2}{(1+f)^2}\nonumber\\
&&+2\frac{f''}{1+f},
\eea
where
\be
\rho_{DE}=6(1+f)\cH f'-3\left(f'\right)^2-3\cH ^2(2+f)f,
\ee
and
\bea
p_{DE}=4\cH \frac{f'}{1+f}-3\frac{\left(f'\right)^2}{(1+f)^2}+2\frac{f''}{1+f},
\eea
respectively. An effective dark energy term,  satisfying the condition $p_{DE}=w\rho_{DE}$, $w={\rm constant}$ can be obtained if the function $f$ satisfies the equation
\bea
&&\frac{2 f''}{1+f}+2 \cH \left[-3 w(1+f)
   +\frac{2}{1+f}\right] f' \nonumber\\
  && +3\left[
   w-\frac{1}{(1+f)^2}\right] \left(f'\right)^2+3 w f (2+f)
   \cH ^2=0.
\eea

The dynamical cosmological dark energy is thus dependent on the scale factor, and on the properties of $\eta$. In the following we will consider only the late time evolution of the Universe, and hence we take $p=0$. To simplify the mathematical formalism we introduce the dimensional time parameter $\tau=\cH _0x^0$, and the normalized Hubble function $h$, defined as $\cH =\cH _0h$, where $\cH_0=H_0/c$, and $H_0$ is the present day value of the Hubble function. Moreover, we denote $u=df/d\tau$. Then the equations describing the cosmological dynamics of the Barthel-Kropina-FLRW cosmological model take the form
\be\label{de1}
\frac{df}{d\tau}=u,
\ee
\be\label{de2}
2\frac{dh}{d\tau}+3h^2=4h\frac{u}{1+f}-3\frac{u^2}{(1+f)^2}+\frac{2}{1+f}\frac{du}{d\tau},
\ee
\bea\label{de3}
&&\frac{2}{1+f}\frac{du}{d\tau}+2h\left[\frac{2}{1+f}-3w(1+f)\right]u\nonumber\\
&&+3\left[w-\frac{1}{(1+f)^2}\right]u^2+3wf(2+f)h^2=0.
\eea

By introducing the critical density $\rho _c=3H_0^2/8\pi G$, and by defining the matter density parameter as $\Omega _m=\rho /\rho _c$, from the first Friedmann equation we obtain
\be
\Omega _m=h^2+\left(\frac{df}{d\tau}\right)^2+(2+f)fh^2-2(1+f)h\frac{df}{d\tau}.
\ee

In terms of the redshift variable $1+z=1/a$, the system of equations (\ref{de1})-(\ref{de3}) become
\be\label{deq1}
-(1+z)h\frac{df}{dz}=u,
\ee
\bea\label{deq2}
-2(1+z)h\frac{dh}{dz}+3h^2&=&4h\frac{u}{1+f}-3\frac{u^2}{(1+f)^2}\nonumber\\
&&-2(1+z)\frac{h}{1+f}\frac{du}{dz},
\eea
\bea\label{deq3}
&&-2(1+z)\frac{h}{1+f}\frac{du}{dz}+2h\left[\frac{2}{1+f}-3w(1+f)\right]u\nonumber\\
&&+3\left[w-\frac{1}{(1+f)^2}\right]u^2+3wf(2+f)h^2=0.
\eea

 For the redshift dependence of the matter density parameter we find
 \be
 \Omega _m=h^2\Bigg[1+(1+z)^2\left(\frac{df}{dz}\right)^2+(2+f)f+2(1+z)(1+f)\frac{df}{dz}\Bigg].
 \ee

 For the deceleration parameter we obtain
 \be
 q=\frac{1}{2}+\frac{3}{2}\frac{\frac{8\pi G}{c^4}\frac{p}{(1+f)^2}-p_{DE}}{\frac{8\pi G}{c^2}\rho +\rho_{DE}}.
 \ee

\begin{figure}[tbp!]
\centering
\includegraphics[width=8cm]{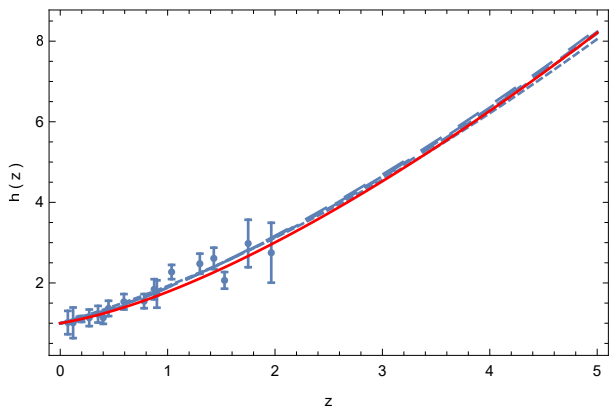}
\caption{Variation of the normalized function $h$ as a function of the redshift $z$ in the Barthel-Kropina-FLRW dark energy model  for different values of the parameter $w$ of the dark energy equation of state:  $w=1.45$ (short dashed curve), $w=1.85$ (dashed curve), $w=2.05$ (long dashed curve), $w=2.55$ (ultra-long dashed curve), and $w=3.05$ (ultra-ultra-long dashed curve), respectively. The red solid line represents the prediction of the $\Lambda$CDM model. The observational data are represented together with their error bars. The initial conditions used to numerically integrate the system of cosmological evolution equations are $f(0)=0.33$, $u(0)=0.51$, and $h(0)=1$, respectively.}
\label{fig4}
\end{figure}

 The variation of the Hubble function, obtained by numerically solving the system of differential equations Eqs.~(\ref{deq1})-(\ref{deq3}), is represented, for different values of the parameter $w$ of the dark energy equation of state, in Fig.~\ref{fig4}.

\begin{figure}[tbp!]
\centering
\includegraphics[width=8cm]{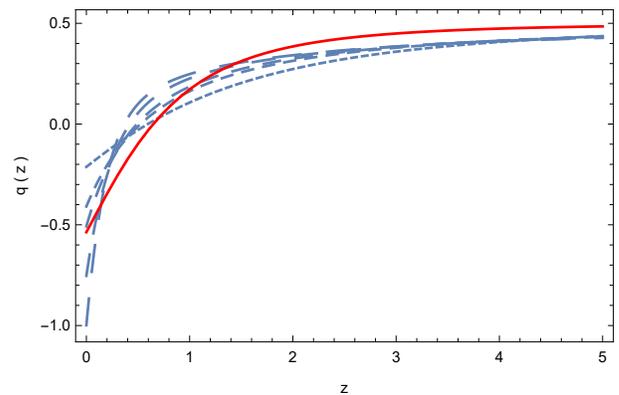}
\caption{Variation of the deceleration parameter $q$ as a function of the redshift $z$ in the Barthel-Kropina-FLRW dark energy model  for different values of the parameter $w$ of the dark energy equation of state:  $w=1.45$ (short dashed curve), $w=1.85$ (dashed curve), $w=2.05$ (long dashed curve), $w=2.55$ (ultra-long dashed curve), and $w=3.05$ (ultra-ultra-long dashed curve), respectively. The red solid line represents the prediction of the $\Lambda$CDM model. The initial conditions used to numerically integrate the system of cosmological evolution equations are $f(0)=0.33$, $u(0)=0.51$, and $h(0)=1$, respectively.}
\label{fig5}
\end{figure}

As one can see from Fig.~\ref{fig4}, the model gives a good description of the observational data, as well as of the $\Lambda$CDM model up to a redshift of at least $z=5$. For higher redshifts some differences with respect to the standard cosmology do appear. The numerical results do not show a significant influence of the variation of the parameter of the dark energy equation of state $w$, with $w=1.85$ predicting almost the same evolution as $w=3.05$, but depend strongly on the initial conditions for $f$ and $u=df/d\tau$. The variation of the deceleration parameter as a function of redshift as predicted by the Barthel-Kropina-FLRW dark energy model is represented in Fig.~\ref{fig5}.

Even that on a qualitative level the behavior of the deceleration parameter of the Barthel-Kropina-FLRW model can reproduce the predictions of the $\Lambda$CDM model, still significant quantitative differences do appear. In the Finslerian approach the evolution of the deceleration parameter is strongly dependent on the parameter $w$ of the dark energy equation of state. At higher redshifts the numerical values of the Barthel-Kropina-FLRW are slightly lower than those obtained from $\Lambda$CDM, and almost independent of $w$, while at low redshifts the behavior of $q$ essentially depends on the numerical value of $w$. The critical redshift $z_{cr}$ indicating the transition from deceleration to acceleration, given by $q\left(z_{cr}\right)=0$, also takes numerical values that are different from the $\Lambda$CDM predictions. The critical redshift is also strongly dependent on the initial conditions at the present time for $f$ and its derivative.

\begin{figure}[tbp!]
\centering
\includegraphics[width=8cm]{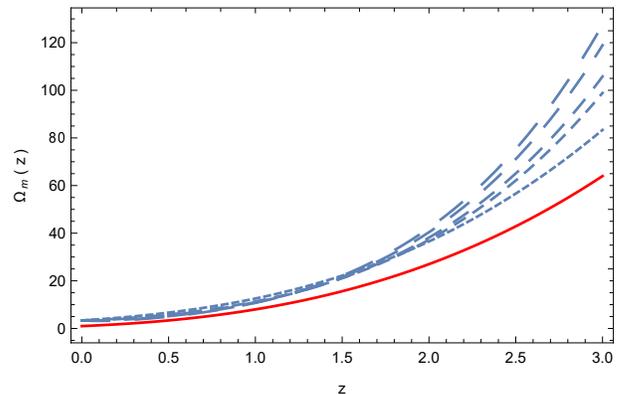}
\caption{Variation of the density parameter of the  matter  $\Omega _m$ as a function of the redshift $z$ in the Barthel-Kropina-FLRW dark energy model  for different values of the parameter $w$ of the dark energy equation of state:  $w=1.45$ (short dashed curve), $w=1.85$ (dashed curve), $w=2.05$ (long dashed curve), $w=2.55$ (ultra-long dashed curve), and $w=3.05$ (ultra-ultra-long dashed curve), respectively. The red solid line represents the prediction of the $\Lambda$CDM model.  The initial conditions used to numerically integrate the system of cosmological evolution equations are $f(0)=0.33$, $u(0)=0.51$, and $h(0)=1$, respectively.}
\label{fig6}
\end{figure}

The density parameter of {\it the total energy-matter content of the Universe}, is represented, as a function of the redshift, in Fig.~\ref{fig6}. In this case there are significant differences, especially at high redshifts,  between the predictions of the Barthel-Kropina-FLRW and the $\Lambda$CDM model, with the Finsler type cosmological model predicting much higher matter densities at higher redshifts. In the present model the matter energy density depends not only on the Hubble function, as in standard general relativity, but also includes the contribution coming from the one-form field $\beta$. Since there is no independent conservation of the matter energy-momentum tensor, the increase of the total matter density at high redshifts can also be interpreted as describing a particle creation process via transfer of energy from the $\eta $ field to ordinary matter, or, equivalently, to the increase in the contribution of the dark energy. On the other hand, in the present model we cannot distinguish between baryonic and dark matter, and dark energy, respectively, and our definition of the density contains all these three components. Hence, $\Omega _m$ includes the contributions of all matter and energy forms of the Universe.

\begin{figure}[tbp!]
\centering
\includegraphics[width=8cm]{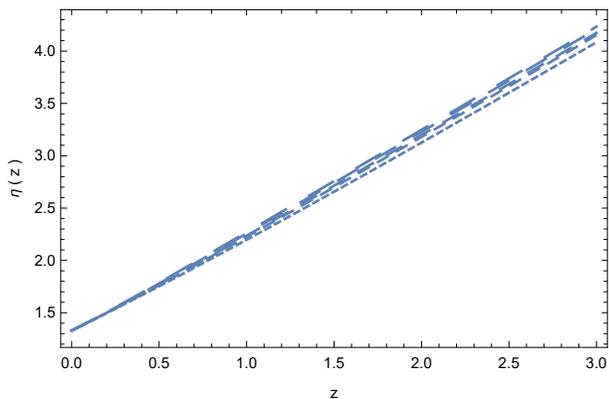}
\caption{Variation of the function $\eta (z)=(1+z)(1+f(z))$ as a function of the redshift $z$ in the Barthel-Kropina-FLRW dark energy model  for different values of the parameter $w$ of the dark energy equation of state:  $w=1.45$ (short dashed curve), $w=1.85$ (dashed curve), $w=2.05$ (long dashed curve), $w=2.55$ (ultra-long dashed curve), and $w=3.05$ (ultra-ultra-long dashed curve), respectively.   The initial conditions used to numerically integrate the system of cosmological evolution equations are $f(0)=0.33$, $u(0)=0.51$, and $h(0)=1$, respectively.}
\label{fig7}
\end{figure}

The variation as a function of the redshift of the function $\eta$ is represented in Fig.~\ref{fig7}. The coefficient of the one-form $\beta$ of the Finsler metric is a monotonically increasing function of $z$ (a monotonically decreasing function of time), and it is a linear function of the redshift. At small redshifts the evolution of $\eta$ is almost independent on the numerical values of $w$, but some small differences do appear at higher redshifts.

\section{Discussions and final remarks}\label{sect5}

In the present paper we have investigated the cosmological implications of a particular Finsler type geometry, in which the fundamental function is obtained as the ratio of a Riemannian metric $\alpha$, and a one form $\beta$, so that $F=\alpha ^2/\beta$. The corresponding geometry is called Kropina geometry, and its properties have been intensively investigated from a mathematical point of view. We have assumed that the physical cosmological metric of the Universe is the FLRW metric, with the help of which the metric $\alpha^2$ is constructed. From a geometric point of view we have adopted the mathematical formalism of the osculating Finsler spaces, in which the Finsler metric $\hat{g}(x,y)$ is localized via the functional relation $y=y(x)$, and thus a Riemann type metric $g(x,y(x))$ is generated. In our study we have considered $A$-osculating Riemannian manifolds $\hat{g}(x,A)$, where $A_I$ are the components of the one form $\beta$.  Moreover, we assume that the Finsler space is an $n$-dimensional  point Finsler space (locally Minkowskian, but generally non Euclidian). The connection of a point Finsler space is the Barthel connection, a connection that depends on the field to which it is applied. For an $(\alpha, \beta)$ metric, the Barthel connection is the Levi-Civita connection of the $A$-Riemannian metric $\hat{g}(x,A)$. From a physical point of view we have postulated that the gravitational phenomena can be described by the $A$-Riemannian metric via the standard Einstein gravitational field equations, in which the Riemannian metric $g(x)$ is substituted by $\hat{g}(x,A)$. Hence, the present approach is based on an extended (but still Riemannian) metric, with the connection and curvature constructed in the standard way.

By adopting as the $\alpha $ metric the standard Friedmann-Lemaitre-Robertson-Walker from, the generalized Friedmann equations can be obtained by a number of straightforward calculations. Despite the apparent complexity of the mathematical model, the generalized Friedmann equations have a very simple mathematical form, and they are obtained in terms of two geometric quantities, the scale factor $a$, and the component $A_0=a\eta$ of the one form $\beta$. The generalized Friedmann equations have the remarkable property of giving in the limit $\eta\rightarrow 1/a$ the standard Friedmann equations of general relativity. Thus, the Barthel-Kropina-FLRW model represents a non-trivial deformation of the standard general relativistic cosmology. In the present work we have performed a systematic study of the cosmological properties of the Finslerian type model, which, despite its close relation with standard general relativistic cosmology also has some very different properties. The generalized Friedmann equations of the Barthel-Kropina-FLRW model do not admit a vacuum solution for $\rho =p=0$, since the first Friedmann equation gives $\eta '=0$, $\eta ={\rm constant}$, with the second field equation identically satisfied. This situation has some similarities with standard general relativity, in which the vacuum Friedmann equations $3H^2=\left(8\pi G/c^2\right)\rho$ and $2\dot{H}+3H^2=-\left(8\pi G/c^4\right)p$ give for $\rho=p=0$ the solution $H=0$, $a={\rm constant}$. Thus the cosmological background solution of the vacuum gives the Minkowski geometry. On the other hand, in the Brathel-Kropina-FLRW model the vacuum solution requires $\eta '=0$, $\eta ={\rm constant}$, a condition that does not fix any background space-time geometry, which remains of the FLRW type, but with arbitrary scale factor $a$.

However, the model admits a de Sitter type solution for the case of a pressureless cosmological fluid, in the presence of a non-zero matter density. The behavior of $\rho$ in this case depends, as one can see from Eq.~(\ref{rhodS}) on an arbitrary integration constant $C_1$. If this constant is taken as $C_1=1$, $\lim_{x^0\rightarrow 0}\rho =\infty$, and therefore the de Sitter evolution begins from a singular state. Finite initial density states are also possible, and in the large time limit the matter density, as well as $\eta$, tend towards some constant values.

In the Barthel-Kropina-FLRW model the matter energy density is generally not conserved, and the energy density and pressure satisfy the balance equation (\ref{bal}). Conservative cosmological models, obtained by imposing the condition of the energy conservation $\rho' +3\cH \left(\rho c^2+p\right)=0$, lead to the standard general relativistic evolution for the scale factor in the presence of matter satisfying a linear barotropic equation of state, $a=\left(x^0\right)^{2/3\gamma}$. Thus, the cosmology of the conservative Barthel-Kropina-FLRW model is identical with the standard general relativistic one.  Hence, accelerated expansion is not possible in the presence of matter satisfying the ordinary matter conservation equations. On the other hand, a large number of accelerating cosmological models can be obtained
with the use of the general energy balance equation (\ref{bal}), once the functional form of $\eta $ is fixed. For the sake of the illustration of the qualitative properties of the model we have considered two particular cosmological scenarios, with $\eta (a)\propto a^{-n}$, and $\eta=\eta (H)$. The first choice of $\eta$ leads to and eternally accelerating Universe, with $q={\rm constant}<0$, while the second choice leads to a plethora of cosmological models whose properties depend on the choice of the function $\eta (h)$. We have considered, and analyzed in detail, a particular class of models obtained by choosing $\eta (H)=a_1H^n+b_1$. The cosmological parameters of this model have been compared with both the observational data and with the predictions of the $\Lambda$CDM model. Even without the use of a proper fitting procedure for the analysis of the observational data, the Barthel-Kropina-FLRW model can give at least an acceptable qualitative description of the observations of the Hubble function, and reproduce the standard $\Lambda$CDM model predictions for $H$. However, significant differences between the Barthel-Kropina-FLRW and the $\Lambda$CDM model do appear in the evolution of the deceleration and matter density parameters. In order to test the validity of the present model a detailed statistical analysis of its predictions with several classes of observational data is necessary.

An effective dark energy model can also be constructed within the framework of the Barthel-Kropina-FLRW geometric theory. Since in the limit $\eta \rightarrow 1/a$ one fully recovers the standard Friedmann equations, one can consider small deviations from the standard Friedmann cosmology by adding a correction term $f$ to $1/a$ in the expression of $\eta$. This procedure allows the reformulation of the generalized Friedmann equations as the standard ones plus some new terms, which are essentially geometric. They can be interpreted as a geometric dark energy, and they contribute effective energy density $\rho _{DE}$ and pressure $p_{DE}$ terms to the standard baryonic thermodynamic quantities. By imposing a linear equation of state for the dark energy terms, $p_{DE}=w\rho _{DE}$, the system of the cosmological equations can be closed, and reformulated as a first order dynamical system in the redshift space. The solutions of the system can be obtained numerically for different numerical values of the parameter $w$ of the dark energy equation of state and for different initial conditions of the function $f$ and of its derivative $f'$. We have performed a detailed comparison of the solutions of the system with the observational data, and the predictions of the $\Lambda$CDM model. The Barthel-Kropina-FLRW model gives a good description of the observational data, and of the $\Lambda$CDM model up to at least a redshift of $z=5$. From the numerical simulations it follows that the numerical values of $w$ have a relatively small influence on the cosmological behavior, which is strongly influenced by the present day values of $f$ and $f'$.

Bouncing cosmological models represent an attractive alternative view of the early Universe, and many such types of models have been proposed to explain the origin of the Universe  (see \cite{new1} for a review).  Bouncing cosmologies have the attractive feature of providing a solution of the singularity problem that generally appear in the standard cosmological models. In the framework of the Finsler and Finsler-like geometries the problem of the bouncing cosmologies was investigated in \cite{new2}.  As a first general result it was shown that in general very special relativity and in Finsler-like gravity on the tangent bundle, the bounce cannot be obtained easily. But in the Finsler-Randers geometry, by adopting for the scale factor the expression $a(t)=a_b\left(1+Bt^2\right)^{1/3}$,  the scalar anisotropy induced by the geometry can satisfy the bounce conditions. Thus, bouncing solutions can be obtained in this class of gravitational theories. In theories constructed by using a nonlinear connection in which a scalar field does appear, with an induced scalar-tensor structure, bouncing solutions can be also obtained. Interestingly enough, in the case of the nonholonomic basis if one imposes to the quantity $N_0(t)$, where $N_{\mu}\equiv \partial _{\phi ^{(\alpha)}}N_{\mu}^{(\alpha)}$, a specific expression in terms of the scalar field and the Hubble function, then this $N_0$ can generate the bouncing scale factor.

Bouncing solutions can also be obtained, at least in principle,  in the Barthel-Kropina cosmological model. By fixing in advance the form of the scale factor so that it has the required bouncing properties, the generalized Friedmann equations (\ref{Fr1}) and (\ref{Fr2}) can be reduced to a (strongly nonlinear) system of differential equations for $\eta$, in which the matter energy density and pressure terms are also present. The investigation of this system can be done only by using numerical (or perturbative) methods. However, in order to obtain bouncing solutions a nonzero matter energy density must be present in the very early Universe.

There are at least three possibilities of the possible theoretical understanding of the vast amount of cosmological data that have radically changed our view on the Universe. The first approach is called the {\it dark components model}, and it generalizes the Einstein gravitational field equations by adding two new terms in the total cosmological energy momentum tensor, which corresponds to dark energy and dark matter, respectively. In this pathway the gravitational dynamics is described by the generalized Einstein equation \cite{HL20}
\be
G_{\mu \nu}=\kappa ^2 T^{\rm bar}_{\mu \nu}+\kappa ^2T^{\rm DM}_{\mu \nu}(\phi, \psi _{\mu},...)+\kappa ^2T^{\rm DE}_{\mu \nu}(\phi, \psi _{\mu},...),
\ee
where  $T^{\rm bar}_{\mu \nu}$, $T^{\rm DM}_{\mu \nu}(\phi, \psi_{\mu},...)$, and $T^{\rm DE}_{\mu \nu}(\phi, \psi_{\mu},...)$ represents the energy-momentum tensors of baryonic matter, dark matter and dark energy, respectively.  Hence, in this extension of general relativity dark energy is a form of matter, an interpretation suggested by the equivalence principle of the theory of relativity that states the equivalence of mass and energy. The energy-momentum tensors of dark energy and dark matter can be realized by using scalar or vector fields.  The simplest dark constituent model can be obtained by assuming that dark energy is a scalar field $\phi$ with a self-interaction potential $V(\phi)$. For reviews and extensive discussions of the dark component models see \cite{PeRa03, Pa03,DE1,DE2,DE3,DE4}).

A second approach to the geometric description of the gravitational interaction is the {\it dark gravity} approach. The dark gravity formalism assumes a purely geometric description of the gravitational phenomena, and it is based on the  modification of the geometric structure of the Einstein field equations, still formulated in a Riemannian geometry. In the dark gravity approach, one can formulate generally the Einstein gravitational field equations as
\be
G_{\mu \nu}=\kappa ^2T_{\mu \nu}^{(mat)}+\kappa ^2 T_{\mu \nu}^{(\rm geom)}\left(g_{\mu \nu}, R, \square R,...\right),
\ee
where $T_{\mu \nu}^{(\rm geom)}\left(g_{\mu \nu}, R, \square R,...\right)$, is a term generating geometrically an effective energy-momentum tensor, constructed from the Riemannian metric.  The geometric term $T_{\mu \nu}^{(\rm geom)}\left(g_{\mu \nu}, R, \square R,...\right)$  can describe dark matter, dark energy, or even both. An interesting example of dark gravity is the $f(R)$ theory \cite{Bu1}, in which the standard Hilbert-Einstein action $S=\int{\left(R/\kappa ^2+L_m\right)\sqrt{-g}d^4x}$ is generalized  a to an action given by $S= \int{\left[f(R)/\kappa^2+L_m\right]\sqrt{-g}d^4x}$, where $f(R)$ is an arbitrary analytical function of the Ricci scalar $R$. For reviews of dark gravity theories, and their applications, see \cite{R1,R2,R3,R4, R5}.

There is a third avenue for the understanding of the present day cosmological data, called {\it the dark coupling} theory. In this approach the standard Hilbert-Einstein Lagrangian density, having a simple additive structure in the geometric and matter Lagrangian,  is replaced by a Lagrangian with a more general algebraic system. In the dark coupling theory  one looks for the maximal extension of the standard Hilbert-Einstein gravitational Lagrangian, and this extension can be obtained by assuming that the gravitational Lagrangian density is an arbitrary analytical function of the curvature scalar $R$, describing the geometric properties of the spacetime, and of the matter Lagrangian $L_m$. Other thermodynamic or geometric parameters can alos be naturally included in specific models. The dark coupling approach leads naturally to the existence of a {\it nonminimal coupling between curvature and matter}.

In the dark coupling approach the Einstein gravitational field equations can be formulated as
\bea
G_{\mu \nu}=\kappa ^2T_{\mu \nu}^{(mat)}+
\kappa ^2 T_{\mu \nu}^{(\rm coup)}\left(g_{\mu \nu},  R, L_m, T, \square R, \square T,... \right),\nonumber\\
\eea
 with the effective energy-momentum tensor of the theory $T_{\mu \nu}^{(\rm coup)}\left(g_{\mu \nu}, R, L_m, T, \square R, \square T,... \right)$ obtained by considering the maximal extension of the Hilbert-Einstein Lagrangian. Moreover, a {\it non-additive curvature-matter algebraic structure} is introduced, describing the couplings between matter and space-time geometry.  The dark coupling theories were considered in \cite{fLm1}, where a gravitational action of the form $S=\int{\left[f_1(R)+\left(1+\lambda f_2(R)\right)L_m\right]\sqrt{-g}d^4x}$ was proposed. This action was generalized in \cite{fLm2} and \cite{fLm2}, leading finally to the $f\left(R,L_m\right)$ gravity theory \cite{fLm3}, in which the gravitational Lagrangian density is given by an arbitrary function of the Ricci scalar and of the matter Lagrangian, $S=\int{f\left(R,L_m\right)\sqrt{-g}d^4x}$. One can also couple curvature and matter via the trace of the matter energy-momentum tensor,  like in the $f(R,T)$ theory, with action given by $S=\int{\left[f\left(R,T\right)+L_m\right]\sqrt{-g}d^4x}$ \cite{fT1}. For a detailed presentation of the theories with curvature-matter coupling see \cite{e8}.

 However, a fourth possibility for the description of the gravitational dynamics and evolution, including the accelerating expansion is also possibly in the framework of the {\it dark geometry} approach, in which one assumes that the true geometry of the nature is beyond the Riemann one, and that the extra terms generated by the non-Riemannian mathematical structures may be responsible for the existence of dark matter and dark energy. One possible dark geometry candidate is Weyl geometry, with conformally invariant gravitational models explaining the present and early dynamics of the Universe \cite{GhHa}. In the present work we have presented another example of a dark geometry, the Finsler type Barthel-Kropina-FLRW geometry, in which an effective dark energy can be generated from the mathematical structures underlying the geometry. The gravitational field equations are postulated as having a similar form with the Einstein equations in Riemann geometry, but with the curvature tensors replaced by their Finslerian counterparts. There is a close relation between the Riemannian cosmological evolution equations, and the Finslerian ones, and this relation allows the natural introduction of a geometric dark energy term in the gravitational formalism.

 Hence, the Barthel-Kropina-FLRW model may represent not only an attractive alternative to the standard $\Lambda$CDM model, but could also open new avenues for the understanding of the complex relation between mathematics and physical reality.

\section*{Acknowledgments}

We would like to thank the anonymous reviewer for comments and suggestions that helped us to significantly improve our manuscript. The work of TH is supported by a grant of the Romanian Ministry of Education and Research, CNCS-UEFISCDI, project number PN-III-P4-ID-PCE-2020-2255 (PNCDI III).

\appendix

\section{Computation of the Barthel-Kropina metric}\label{app1}

\subsection{Method 1}

We recall the formula for the fundamental tensor of this Kropina metric in \cite{YS}
\bea\label{Kropina Hessian}
\hat g_{IJ}(x,y)&=&\frac{2\alpha^2}{\beta^2}\epsilon g_{IJ}(x)+\frac{3\alpha^4}{\beta^4}b_Ib_J-\frac{4\alpha^2}{\beta^3}(\widetilde{g}_{0I}b_J+\widetilde{g}_{0J}b_I)\nonumber\\
&&+\frac{4}{\beta^2}\widetilde{g}_{0I}\widetilde{g}_{0J},
\eea
where $\widetilde{g}_{0I}:=\epsilon g_{IJ}y^J$ (see \cite{YS}). {\it Pay attention that the index zero in the} $\widetilde{g}$ {\it terms means contraction by} $y^I$.

We will compute with the help of the above formula the components of the metric tensor under the conditions specified below.

{\bf Step 1.} Assume

\begin{enumerate}[(i)]
\item $\epsilon=1$;
\item $\left(g_{IJ}(x)\right) =\begin{pmatrix}
1 & 0 & 0 & 0 \\
0 & -a^2(x^0) & 0 & 0 \\
0 & 0 & -a^2(x^0) & 0\\
0 & 0 & 0 & -a^2(x^0)
\end{pmatrix};$
\item $(A_I(x))=\begin{pmatrix}
    A_0\\0\\0\\0
  \end{pmatrix}=\begin{pmatrix}
    a(x^0)\eta(x^0)\\0\\0\\0
  \end{pmatrix}.
  $
\end{enumerate}

Under these assumptions formula \eqref{Kropina Hessian} gives
\begin{equation}
\left(\hat{g}_{IJ}(x,y)\right) =\begin{pmatrix}
\hat g_{00}(x,y) & \hat g_{0i}(x,y)) \\
(\hat g_{0i}(x,y)) & (\hat g_{ij}(x,y))
\end{pmatrix},
  \end{equation}
  with the components
  \begin{eqnarray*}
      \hat{g}_{00}(x,y)&=&\frac{2\alpha^2}{\beta^2}+\frac{3\alpha^4}{\beta^4}a^2\eta^2-\frac{8\alpha^2}{\beta^3}a\eta y^0+\frac{4}{\beta^2}(y^0)^2,\\
      \hat{g}_{0i}(x,y)&=&\frac{4a^2y^i}{\beta^3}\left(\alpha^2b_0-\beta y^0 \right), \\
      \hat{g}_{ij}(x,y)&=&-\frac{2a^2}{\beta^2}\left[\delta_{ij}\alpha^2-2a^2y^iy^j
        \right].
    \end{eqnarray*}

Observe that in the right hand side, the indices do not obey strictly the covariant writing form. However, this is not a problem now, since they indicate merely the position than the summation.

    {\bf Step 2.} We evaluate this matrix for arbitrary $x$ and the specific direction
    \begin{equation}
y=(y^I)=A=\begin{pmatrix}
    A_0\\0\\0\\0
  \end{pmatrix}=\begin{pmatrix}
    a(x^0)\eta(x^0)\\0\\0\\0
  \end{pmatrix}.
      \end{equation}

      Then the Kropina fundamental tensor reads

  \bea
&&\left(\hat g_{IJ}(x,y)|_{y=A}\right)\nonumber\\
      &&    =\begin{pmatrix}
\dfrac{1}{a^2(x^0)\eta^2(x^0)} & 0 & 0 & 0 \\
0 & \dfrac{-2}{\eta^2(x^0)} & 0 & 0 \\
0 & 0 & \dfrac{-2}{\eta^2(x^0)} & 0\\
0 & 0 & 0 & \dfrac{-2}{\eta^2(x^0)}
\end{pmatrix}.
\eea

\subsection{Method 2}

Let $\alpha=\sqrt{\epsilon g_{IJ}(x)y^Iy^J}$ and $\beta=A_I(x)y^I$, and
\be
F=\alpha\phi(s),\ s=\frac{\beta}{\alpha},
\ee
The Hessian $\hat{g}$
\be
\hat{g}_{IJ}=\epsilon \rho g_{IJ}+\rho_0b_Ib_J+\rho_1\left(b_I\alpha_J+b_J\alpha_I\right)-s\rho_1\alpha_I\alpha_J,
\ee
where $\alpha_I:=\alpha_{y^I}$ and
$$
\rho=\phi^2-s\phi\phi', \ \rho_0=\phi\phi''+\phi'\phi',\ \rho_1=-s(\phi\phi''+\phi'\phi')+\phi\phi'.
$$

We obtain first
\begin{equation*}
\begin{split}
\alpha_I&=\alpha_{y^I}=\frac{\partial \alpha}{\partial y^I}=\frac{\partial}{\partial y^I}\sqrt{\epsilon g_{MN}y^My^N}\\
&=\frac{1}{2\alpha}\epsilon g_{MN}y^M\delta^N_I+\frac{1}{2\alpha}\epsilon g_{MN}y^N\delta^M_I\\
&=\frac{1}{2\alpha}\epsilon g_{MI}y^M+\frac{1}{2\alpha}\epsilon g_{IN}y^N\\
&=\frac{1}{2\alpha}\epsilon y_I+\frac{1}{2\alpha}\epsilon y_I=\epsilon\frac{y_I}{\alpha}.
\end{split}
\end{equation*}

We consider now $y^i=A^i$, then
\begin{equation*}
\begin{split}
\alpha&=\sqrt{\epsilon g_{IJ}(x)A^IA^J}=\sqrt{\epsilon A_IA^J}=\sqrt{\epsilon A^2}\\
\beta&=A_IA^I=A^2=\epsilon\alpha^2,\\
s&=\frac{\beta}{\alpha}=\frac{\epsilon\alpha^2}{\alpha}=\epsilon\alpha,
\alpha_I=\epsilon\frac{A_I}{\alpha}.
\end{split}
\end{equation*}

Therefore
\begin{equation*}
\begin{split}
\hat{g}_{IJ}(x,A)&=\epsilon \rho g_{IJ}+\rho_0A_IA_J+\rho_1\left(A_I\epsilon\frac{A_J}{\alpha}+A_J\epsilon\frac{A_I}{\alpha}\right)\\
&-\epsilon\alpha\rho_1\epsilon\frac{A_I}{\alpha}\epsilon\frac{A_J}{\alpha}\\
&=\epsilon \rho g_{IJ}+\rho_0A_IA_J+2\epsilon\rho_1\frac{A_IA_J}{\alpha}-\epsilon\rho_1\frac{A_IA_J}{\alpha}\\
&=\epsilon \rho g_{IJ}+(\alpha\rho_0+\epsilon\rho_1)\frac{A_IA_J}{\alpha}.
\end{split}
\end{equation*}

For the Kropina metric we let $\phi(s)=\dfrac{1}{s}$, it follows that
\begin{equation*}
\begin{split}
\phi'(s)=\frac{-1}{s^2},
\phi''(s)=\frac{2}{s^3}.
\end{split}
\end{equation*}

Therefore
\begin{equation*}
\begin{split}
\rho=\frac{2}{s^2},\rho_0=\frac{3}{s^4},\rho_1=-\frac{4}{s^3}.
\end{split}
\end{equation*}

Substitute $s=\epsilon\alpha$, we obtain
$$
\rho=\frac{2}{\alpha^2},\ \rho_0=\frac{3}{\alpha^4},\ \rho_1=\frac{-4}{\epsilon\alpha^3}=\frac{-4\epsilon}{\alpha^3}.
$$

Next, we substitute to $\hat{g}(x,A)$, thus obtaining
\bea\label{Kropm}
\hat{g}(x,A)&=&\epsilon \rho g_{IJ}+(\alpha\rho_0+\epsilon\rho_1)\frac{A_IA_J}{\alpha}\nonumber\\
&=&\frac{2\epsilon}{\alpha^2}g_{IJ}+\left(\frac{3}{\alpha^3}-\frac{4}{\alpha^3}\right)\frac{A_IA_J}{\alpha}\nonumber\\
&=&\frac{2\epsilon}{\alpha^2}g_{IJ}-\frac{1}{\alpha^3}\frac{A_IA_J}{\alpha}\nonumber\\
&=&\frac{1}{\alpha^2}\left(2\epsilon g_{IJ}-\frac{A_IA_J}{\alpha^2}\right).
\eea

\subsubsection{Computation of $\hat{g}(x,A)$}

Let's now consider the formula (\ref{Kropm}) of $\hat{g}(x,A)$ together with the assumptions
\begin{enumerate}[(i)]
\item $\epsilon=1$;
\item $\left(A_I\right)=\left(a\left(x^0\right)\eta\left(x^0\right),0,0,0\right)=\left(A^I\right)$;
\item $\left(g_{IJ}\right)=\begin{pmatrix}
1 & 0 & 0 & 0 \\
0 & -a^2(x^0) & 0 & 0 \\
0 & 0 & -a^2(x^0) & 0\\
0 & 0 & 0 & -a^2(x^0)
\end{pmatrix};$
\item $\alpha|_{y=A(x)}=a(x^0)\eta(x^0)$;
\item $\beta|_{y=A(x)}=[a(x^0)\eta(x^0)]^2$.
\end{enumerate}

Then we immediately obtain
\begin{equation*}
\begin{split}
\hat{g}_{00}&=\frac{1}{\alpha^2}\left(2g_{00}-\frac{A_0^2}{\alpha^2}\right)
=\frac{1}{a^2\eta^2}\left(2(1)-\frac{(a\eta)^2}{(a\eta)^2}\right)\\
&=\frac{1}{a^2\eta^2},\\
\end{split}
\end{equation*}
\begin{eqnarray*}
\hat{g}_{ij}&=&\frac{1}{\alpha^2}\left(2g_{ij}-\frac{A_iA_j}{\alpha^2}\right)
=\frac{1}{\alpha^2}\left(2g_{ii}-\frac{A_i^2}{\alpha^2}\right)\delta_{ij}\nonumber\\
&=&\frac{1}{a^2\eta^2}\left(2(-a^2)\right)\delta_{ij}
=-\frac{2}{\eta^2}\delta_{ij}.
\end{eqnarray*}

Hence the non-vanishing component of $\hat{g}_{IJ}$ are
\begin{equation*}
\hat{g}_{IJ}=\begin{cases}
\hat{g}_{00}&=\dfrac{1}{a^2(x^0)\eta^2(x^0)},\vspace{0.5cm}\\
\hat{g}_{ij}&=\dfrac{-2}{\eta^2(x^0)}\delta_{ij}.
\end{cases}
\end{equation*}

\section{Computation of the Christoffel symbols}\label{app2}

Recall the formula of Christoffel symbols
\be
\hat{\gamma}^A_{BC}=\frac{1}{2}g^{AD}\left(\frac{\partial g_{BD}}{\partial x^C}+\frac{\partial g_{CD}}{\partial x^B}-\frac{\partial g_{BC}}{\partial x^D}\right),
\ee
where
$$
\hat{g}^{IJ}=\begin{pmatrix}
a^2\eta^2 & 0 & 0 & 0 \\
0 & \dfrac{-\eta^2}{2} &0 & 0\\
0 & 0 &  \dfrac{-\eta^2}{2} & 0\\
0 & 0 & 0 &  \dfrac{-\eta^2}{2}
\end{pmatrix}.
$$

We begin by computing the derivatives of $\hat{g}_{IJ}$, that is
\begin{equation*}
\begin{split}
\frac{\partial \hat{g}_{00}}{\partial x^0}&=\frac{\partial}{\partial x^0}\left(\frac{1}{a^2(x^0)\eta^2(x^0)}\right)
=\frac{-1}{a^4\eta^4}(a^2(2\eta\eta')+\eta^2(2aa'))\\
&=-\frac{2(a\eta'+a'\eta)}{a^3\eta^3},\\
\frac{\partial \hat{g}_{ij}}{\partial x^0}&=\frac{\partial}{\partial x^0}\frac{-2}{\eta^2(x^0)}\delta_{ij}
=\frac{4\eta'}{\eta^3}\delta_{ij}.
\end{split}
\end{equation*}

Next, we compute the Christoffel symbols
\begin{equation*}
\begin{split}
\hat{\gamma}^0_{00}&=\frac{1}{2}g^{00}\left(\frac{\partial g_{00}}{\partial x^0}+\frac{\partial g_{00}}{\partial x^0}-\frac{\partial g_{00}}{\partial x^0}\right)\\
&=\frac{1}{2}g^{00}\left(\frac{\partial g_{00}}{\partial x^0}\right)
=\frac{1}{2}a^2\eta^2\left[-\frac{2(a\eta'+a'\eta)}{a^3\eta^3}\right]\\
&=-\frac{(a\eta'+a'\eta)}{a\eta}=-\frac{(\eta\cH+\eta')}{\eta},\\
\end{split}
\end{equation*}
\begin{equation*}
\begin{split}
\hat{\gamma}^0_{ij}&=\frac{1}{2}g^{0D}\left(\frac{\partial g_{iD}}{\partial x^j}+\frac{\partial g_{jD}}{\partial x^i}-\frac{\partial g_{ij}}{\partial x^D}\right)\\
&=\frac{1}{2}g^{00}\left(\frac{\partial g_{i0}}{\partial x^j}+\frac{\partial g_{j0}}{\partial x^i}-\frac{\partial g_{ij}}{\partial x^0}\right)\\
&=\frac{1}{2}g^{00}\left(-\frac{\partial g_{ij}}{\partial x^0}\right)
=\frac{1}{2}a^2\eta^2\left(-\frac{4\eta'}{\eta^3}\right)\delta_{ij}=\frac{-2a^2\eta'}{\eta}\delta_{ij},\\
\end{split}
\end{equation*}
\begin{equation*}
\begin{split}
\hat{\gamma}^i_{0j}&=\frac{1}{2}g^{iD}\left(\frac{\partial g_{0D}}{\partial x^j}+\frac{\partial g_{jD}}{\partial x^0}-\frac{\partial g_{0j}}{\partial x^D}\right)\\
&=\frac{1}{2}g^{is}\left(\frac{\partial g_{js}}{\partial x^0}\right)
=\frac{1}{2}g^{ij}\left(\frac{\partial g_{ji}}{\partial x^0}\right)\\
&=\frac{1}{2}\left(\frac{-\eta^2}{2}\right)\left(\frac{4\eta'}{\eta^3}\right)\delta_{ij}=\frac{-\eta'}{\eta}\delta_{ij},
\end{split}
\end{equation*}
where $\cH=\frac{a'}{a}$.

\section{Computation of the Ricci tensor}\label{app3}

We start by computing the derivatives of the non-vanishing Christoffel symbols,
\begin{equation*}
\begin{split}
&\frac{\partial}{\partial x^0}\hat{\gamma}^0_{00}=\frac{\partial}{\partial x^0}\left(\frac{-(a\eta'+a'\eta)}{a\eta}\right)\\
&=-\frac{a\eta(a\eta''+a'\eta'+a'\eta'+a''\eta)-(a\eta'+a'\eta)(a\eta'+a'\eta)}{a^2\eta^2}\\
&=-\frac{a^2\eta\eta''+2aa'\eta\eta'+aa''\eta^2-a^2(\eta')^2-2aa'\eta\eta'-(a')^2\eta^2}{a^2\eta^2}\\
&=\frac{a^2(\eta')^2+(a')^2\eta^2-a^2\eta\eta''-aa''\eta^2}{a^2\eta^2},\\
&\frac{\partial}{\partial x^0}\hat{\gamma}^0_{ij}=\frac{\partial}{\partial x^0}\left(\frac{-2a^2\eta'}{\eta}\delta_{ij}\right)\\
&=-2\frac{\eta(a^2\eta''+2aa'\eta')-a^2(\eta')^2}{\eta^2}\delta_{ij}\\
&=\frac{2a^2(\eta')^2-2a^2\eta\eta''-4aa'\eta\eta'}{\eta^2}\delta_{ij},\\
&\frac{\partial}{\partial x^0}\hat{\gamma}^i_{0j}=\left(\frac{-\eta'}{\eta}\delta_{ij}\right)
=-\frac{\eta\eta''-(\eta')^2}{\eta^2}\delta_{ij}
=\frac{(\eta')^2-\eta\eta''}{\eta^2}\delta_{ij}.
\end{split}
\end{equation*}

Next, we compute the Ricci tensor. Recall the formula of Ricci tensor, that is
\be
\hat{R}_{BD}=\sum_A\left[\frac{\partial\hat{\gamma}^A_{BD}}{\partial x^A}-\frac{\partial \hat{\gamma}^A_{BA}}{\partial x^D}+\sum_E\left(\hat{\gamma}^E_{BD}\hat{\gamma}^A_{EA}-\hat{\gamma}^E_{BA}\hat{\gamma}^A_{ED}\right)\right].
\ee

Thus we obtain
\begin{equation*}
\begin{split}
\hat{R}_{00}&=\sum_A\left[\frac{\partial\hat{\gamma}^A_{00}}{\partial x^A}-\frac{\partial \hat{\gamma}^A_{0A}}{\partial x^0}+\sum_E\left(\hat{\gamma}^E_{00}\hat{\gamma}^A_{EA}-\hat{\gamma}^E_{0A}\hat{\gamma}^A_{E0}\right)\right]\\
&=\sum_A\Big[\frac{\partial\hat{\gamma}^A_{00}}{\partial x^A}-\frac{\partial \hat{\gamma}^A_{0A}}{\partial x^0}+\hat{\gamma}^0_{00}\hat{\gamma}^A_{0A}-\hat{\gamma}^0_{0A}\hat{\gamma}^A_{00}\\
&+3(\hat{\gamma}^s_{00}\hat{\gamma}^A_{sA}-\hat{\gamma}^s_{0A}\hat{\gamma}^A_{s0})\Big]\\
&=\sum_A\left[\frac{\partial\hat{\gamma}^A_{00}}{\partial x^A}-\frac{\partial \hat{\gamma}^A_{0A}}{\partial x^0}+\hat{\gamma}^0_{00}\hat{\gamma}^A_{0A}-\hat{\gamma}^0_{0A}\hat{\gamma}^A_{00}-3\hat{\gamma}^s_{0A}\hat{\gamma}^A_{s0}\right]\\
&=\frac{\partial\hat{\gamma}^0_{00}}{\partial x^0}-\frac{\partial \hat{\gamma}^0_{00}}{\partial x^0}+\hat{\gamma}^0_{00}\hat{\gamma}^0_{00}-\hat{\gamma}^0_{00}\hat{\gamma}^0_{00}-3\hat{\gamma}^s_{00}\hat{\gamma}^0_{s0}\\
&+3\left[\frac{\partial\hat{\gamma}^k_{00}}{\partial x^k}-\frac{\partial \hat{\gamma}^k_{0k}}{\partial x^0}+\hat{\gamma}^0_{00}\hat{\gamma}^k_{0k}-\hat{\gamma}^0_{0k}\hat{\gamma}^k_{00}\right]-3\hat{\gamma}^s_{0k}\hat{\gamma}^k_{s0}\\
&=3\left[-\frac{\partial \hat{\gamma}^k_{0k}}{\partial x^0}+\hat{\gamma}^0_{00}\hat{\gamma}^k_{0k}\right]-3\hat{\gamma}^s_{0k}\hat{\gamma}^k_{s0}\\
&=3\Bigg[-\frac{(\eta')^2-\eta\eta''}{\eta^2}+\left(\frac{(a\eta'+a'\eta)}{a\eta}\right)\left(\frac{\eta'}{\eta}\right)
-\left(\frac{-\eta'}{\eta}\right)^2\Bigg]\\
&=\frac{3}{\eta^2}\left[-(\eta')^2+\eta\eta''+(\eta')^2+\frac{a'\eta\eta'}{a}-(\eta')^2\right]\\
&=\frac{3}{a\eta^2}\left[a\eta\eta''+a'\eta\eta'-a(\eta')^2\right]\\
&=\frac{3}{\eta^2}\left[\eta\eta''+\eta\eta'\cH-(\eta')^2\right],
\end{split}
\end{equation*}
and
\begin{equation*}
\begin{split}
\hat{R}_{ij}&=\sum_A\left[\frac{\partial\hat{\gamma}^A_{ij}}{\partial x^A}-\frac{\partial \hat{\gamma}^A_{iA}}{\partial x^j}+\sum_E\left(\hat{\gamma}^E_{ij}\hat{\gamma}^A_{EA}-\hat{\gamma}^E_{iA}\hat{\gamma}^A_{Ej}\right)\right]\\
&=\sum_A\left[\frac{\partial\hat{\gamma}^A_{BD}}{\partial x^A}+\hat{\gamma}^0_{ij}\hat{\gamma}^A_{0A}-\hat{\gamma}^0_{iA}\hat{\gamma}^A_{0j}+3\left(\hat{\gamma}^s_{ij}\hat{\gamma}^A_{sA}-\hat{\gamma}^s_{iA}\hat{\gamma}^A_{sj}\right)\right]\\
&=\sum_A\left[\frac{\partial\hat{\gamma}^A_{BD}}{\partial x^A}+\hat{\gamma}^0_{ij}\hat{\gamma}^A_{0A}-\hat{\gamma}^0_{iA}\hat{\gamma}^A_{0j}-\hat{\gamma}^s_{iA}\hat{\gamma}^A_{sj}\right]\\
&=\frac{\partial\hat{\gamma}^0_{ij}}{\partial x^0}+\hat{\gamma}^0_{ij}\hat{\gamma}^0_{00}-\hat{\gamma}^0_{i0}\hat{\gamma}^0_{0j}-\hat{\gamma}^s_{i0}\hat{\gamma}^0_{sj}\\
&+3\left[\frac{\partial\hat{\gamma}^k_{iD}}{\partial x^k}+\hat{\gamma}^0_{ij}\hat{\gamma}^k_{0k}-\hat{\gamma}^0_{ik}\hat{\gamma}^k_{0j}-\hat{\gamma}^s_{ik}\hat{\gamma}^k_{sj}\right]\\
&=\frac{\partial\hat{\gamma}^0_{ij}}{\partial x^0}+\hat{\gamma}^0_{ij}\hat{\gamma}^0_{00}+\hat{\gamma}^s_{i0}\hat{\gamma}^0_{sj}\\
&=\frac{2a^2(\eta')^2-2a^2\eta\eta''-4aa'\eta\eta'}{\eta^2}\\
&+\left(\frac{-2a^2\eta'}{\eta}\right)\left(\frac{-(a\eta'+a'\eta)}{a\eta}\right)
+\left(\frac{-\eta'}{\eta}\right)\left(\frac{-2a^2\eta'}{\eta}\right)\\
&=\frac{2a}{\eta^2}\left(3a(\eta')^2-a\eta\eta''-a'\eta\eta'\right)\delta_{ij}\\
&=\frac{2a^2}{\eta^2}\left(3(\eta')^2-\eta\eta''-\eta\eta'\cH\right)\delta_{ij}.
\end{split}
\end{equation*}

\section{Ricci scalar}\label{app4}

In this Appendix, we will compute the Ricci scalar,
\be
\hat{R}=\hat{g}^{IJ}\hat{R}_{
IJ}.
\ee

We obtain
\begin{equation*}
\begin{split}
\hat{R}&=\hat{g}^{00}\hat{R}_{00}+3\hat{g}^{ii}\hat{R}_{ii}\\
&=a^2\eta^2\left(\frac{3}{a\eta^2}\left[a\eta\eta''+a'\eta\eta'-a(\eta')^2\right]\right)\\
&+3\left(\frac{-\eta^2}{2}\right)\left[\frac{-2a}{\eta^2}\left(-3a(\eta')^2+a\eta\eta''+a'\eta\eta'\right)\right]\\
&=3a^2\eta\eta''+3aa'\eta\eta'-3a^2(\eta')^2\\
&-9a^2(\eta')^2+3a^2\eta\eta''+3aa'\eta\eta'\\
&=6a^2\eta\eta''+6aa'\eta\eta'-12a^2(\eta')^2\\
&=6a(a\eta\eta''+a'\eta\eta'-2a(\eta')^2)\\
&=6a^2\left[\eta\eta''+\cH\eta\eta'-2(\eta')^2\right].
\end{split}
\end{equation*}

\section{The generalized Friedmann equations}\label{app5}

Recall the formula of the Einstein tensor,
\be
\hat{G}_{IJ}=\hat{R}_{IJ}-\frac{1}{2}\hat{R}\hat{g}_{IJ}.
\ee

It follows that
\begin{equation*}
\begin{split}
\hat{G}_{00}&=\hat{R}_{00}-\frac{1}{2}\hat{R}\hat{g}_{00}
=\frac{3}{a\eta^2}\left[a\eta\eta''+a'\eta\eta'-a(\eta')^2\right]\\
&-\frac{1}{2}\left(6a(a\eta\eta''+a'\eta\eta'-2a(\eta')^2)\right)\frac{1}{a^2\eta^2}\\
&=\frac{3}{a\eta^2}\left[a\eta\eta''+a'\eta\eta'-a(\eta')^2-a\eta\eta''-a'\eta\eta'+2a(\eta')^2\right]\\
&=\frac{3(\eta')^2}{\eta^2},
\end{split}
\end{equation*}
and
\begin{equation*}
\begin{split}
\hat{G}_{ii}&=\hat{R}_{ii}-\frac{1}{2}\hat{R}\hat{g}_{ii}
=\frac{2a}{\eta^2}\left(3a(\eta')^2-a\eta\eta''-a'\eta\eta'\right)\\
&-\frac{1}{2}\left(6a(a\eta\eta''+a'\eta\eta'-2a(\eta')^2)\right)\left(\frac{-2}{\eta^2}\right)\\
&=\frac{2a}{\eta^2}\left[3a(\eta')^2-a\eta\eta''-a'\eta\eta'+3a\eta\eta''+3a'\eta\eta'-6a(\eta')^2\right]\\
&=\frac{2a}{\eta^2}\Bigg[-3a(\eta')^2+2a\eta\eta''
+2a'\eta\eta'\Bigg]\\
&=\frac{2a^2}{\eta^2}\left[-3(\eta')^2+2\eta\eta''+2\cH\eta\eta'\right].
\end{split}
\end{equation*}
The rest of the components of the Einstein tensor vanish.

\end{document}